\documentclass[11pt]{article}
\usepackage{fancyhdr}

\usepackage{graphicx}
\usepackage{latexsym}
\usepackage{appendix}
\usepackage{srcltx}
\def\Tr{{\rm Tr}}

\def\R{{\bf R}}

\def\CE{{\cal E}}

\def\CI{{\cal I}}
\def\CL{{\cal L}}

\def\CO{{\cal O}}

\def\centeron#1#2{{\setbox0=\hbox{#1}\setbox1=\hbox{#2}\ifdim
   \wd1>\wd0\kern.48\wd1\kern-.48\wd0\fi
   \copy0\kern-.48\wd0\kern-.48\wd1\copy1\ifdim\wd0>\wd1
   \kern.48\wd0\kern-.48\wd1\fi}}

\newcommand{\beq}{\begin{equation}}
\newcommand{\eeq}{\end{equation}}
\newcommand{\bea}{\begin{eqnarray}}
\newcommand{\eea}{\end{eqnarray}}
\newcommand{\ba}{\begin{array}}
\newcommand{\ea}{\end{array}}

\newcommand{\p}{\partial}
\newcommand{\nn}{\nonumber}

\newcommand{\half}{\frac{1}{2}}

\newcommand{\bfalpha}{{\mbox{\boldmath $\alpha$}}}
\newcommand{\bfbeta}{{\mbox{\boldmath $\beta$}}}
\newcommand{\bfgamma}{{\mbox{\boldmath $\gamma$}}}

\newcommand{\bftau}{{\mbox{\boldmath $\tau$}}}
\newcommand{\bfeta}{{\mbox{\boldmath $\eta$}}}
\newcommand{\bfxi}{{\mbox{\boldmath $\xi$}}}
\newcommand{\bfkappa}{{\mbox{\boldmath $\kappa$}}}
\newcommand{\bfepsilon}{{\mbox{\boldmath $\epsilon$}}}

\begin{document}

\hskip3cm

 \hskip12cm{CQUeST-2010-0392}
\vskip3cm

\begin{center}
 \LARGE \bf  Mass and Angular momentum of Black Holes\\
in  New Massive Gravity
\end{center}

\vskip2cm

\centerline{\Large Soonkeon Nam\footnote{nam@khu.ac.kr}\,,
~~Jong-Dae Park\footnote{jdpark@khu.ac.kr}\,, ~~Sang-Heon
Yi\footnote{shyi@sogang.ac.kr}}

\hskip2cm

\begin{quote}
Department of Physics and Research Institute of Basic Science, Kyung
Hee University, Seoul 130-701, Korea$^{1,2}$

Center for Quantum Spacetime, Sogang University, Seoul 121-741,
Korea$^3$
\end{quote}

\hskip2cm

\vskip2cm

\centerline{\bf Abstract} We obtain  mass and angular momentum of
black holes as conserved charges in  three dimensional new massive
gravity, after presenting the explicit expression for the potential
of the conserved charges.   This confirms the   expression of those
charges obtained in several ways, in particular through AdS/CFT
correspondence,  and  shows us  that the first law of black hole
thermodynamics is valid in these black holes.  We also comment about
conserved charges  in new type black holes with the emphasis on  the
AdS/CFT correspondence as guiding principle. \thispagestyle{empty}
\renewcommand{\thefootnote}{\arabic{footnote}}
\setcounter{footnote}{0}
\newpage

\section{Introduction}
Three dimensional pure Einstein gravity is non-dynamical  and seems
not so relevant to our dynamical four dimensional gravity. Despite
of this apparent irrelevance,   three dimensional gravity becomes an
interesting arena toward understanding some aspects of gravity and
testing various ideas about black holes, since it may contain black
holes and have a dual description through $AdS/CFT$ correspondence.
Furthermore, when additional terms such as higher curvature terms
are added, its physical contents become richer and it has some
interesting aspects in the viewpoint of $AdS/CFT$ correspondence.
The higher curvature terms in three dimension make gravitons
propagate and open a new window of the $AdS/CFT$ correspondence.

Gravitational Chern-Simons term is one of interesting correction
terms, for which it has long been  known that three dimensional pure
gravity with this term~\cite{Deser:1981wh}\cite{Deser:1982} leads to
massive gravitons, allows black hole solutions, and  also has a
string theory embedding~\cite{Clement:1994gr}$\sim$\cite{Nutku:1993}
. This three dimensional gravity with Chern-Simons term has been
called topologically massive gravity(TMG).  This TMG has also been
investigated in the viewpoint of $AdS/CFT$
correspondence(See~\cite{skenderis:2009}$\sim$\cite{Solodukhin:2001}).
Recently, another type of three dimensional massive gravity theory
with a specific combination of higher curvature terms has been
explored (see for a
review~\cite{Bergshoeff:2009hq}$\sim$\cite{Bergshoeff:2009}), which
is called new massive gravity(NMG). The simplest version of this
theory contains a specific combination of  scalar curvature square
and Ricci tensor square, and it preserves parity symmetry contrary
to TMG. It has also been shown that this NMG  allows various black
holes~\cite{Bergshoeff:2009aq}$\sim$\cite{Giribet:2009qz} and also
has another realization of the $AdS/CFT$ correspondence.

Though there are several studies on black holes in NMG, far less is
known than those in TMG.  Mass and angular momentum of black holes
are not obtained as conserved charges and the so-called new type
black holes found in NMG are not so well understood, yet. In this
paper, we obatin mass and angular momentum as conserved charges on
various black holes and then resolve some issues on new type black
holes.

There are several methods to define mass and angular momentum of
black holes for higher curvature theories, for example as in
~\cite{Abbott:82}$\sim$\cite{Nam:2010dd}. One of the easiest way to
obtain those in the case of asymptotically $AdS$ space is to resort
to AdS/CFT correspondence. Specifically, in the three dimensional
case the mass and angular momentum of asymptotically $AdS$ black
holes can be obtained by
\beq M = E_L +E_R\,, \qquad J = L(E_L-E_R)\,,  \label{AsymAdS}\eeq
where $E_L, E_R$ are left and right energies of the dual CFT and $L$
is the $AdS$ radius. In terms of the central charges of dual two
dimensional CFTs,  these energies can be related to the left and
right black hole temperatures as $E_L = \pi^2L/6\cdot c_L T_L^2$,
$E_R = \pi^2L/6\cdot c_R T_R^2$, and central charges of the
conjectural dual CFTs can also be obtained from black hole solutions
through the Cardy formula.  Therefore, the above formulae mean that
mass and angular momentum can be obtained purely from gravity data
without CFT.  For warped $AdS$ black holes, those should be read as
\beq
  M  = \sqrt{\frac{3\beta^4}{4\beta^2-1}}\sqrt{\frac{2c_L E_L}{3L}} \,, \qquad
  J = L (E_L-E_R)\,,
\eeq
where $\beta$ is the warp factor.

These formulae for obtaining mass and angular momentum through
$AdS/CFT$ correspondence are consistent with the first law of black
holes thermodynamics by construction. These formulae also hold in
TMG for the same kind black holes, and can be shown to be identical
with those obtained as conserved charges.  One of our main results
is to verify these formulae in NMG by obtaining mass and angular
momentum of black holes as conserved charges in the purely gravity
context, which can also be understood as verification of the first
law of black hole thermodynamics for these black holes in NMG.

In NMG  there are new black hole solutions, so-called {\it new type
black holes}, in addition to the above mentioned (warped) $AdS$
black holes~\cite{Bergshoeff:2009aq}. Though these solutions are
obtained analytically, a deep understanding of their physical
properties is still lacking. One of motivations to study conserved
charges of black holes  in this paper is to improve this situation
in these black holes, keeping in mind  $AdS/CFT$ correspondence as
the guiding principle.

The asymptotic of new type black holes in NMG is $AdS$ space, which
is identical to the asymptotic of BTZ black holes. Therefore,
according to the standard $AdS/CFT$ correspondence,  new type black
holes and BTZ black holes should correspnd to the related dual CFTs,
while the dual CFTs need not be completely identical.  In
particular, the central charges obtained from new type black holes
and BTZ black holes should be matched for a suitable adjustment of
parameters which is necessary for the existence of new type black
holes in the gravity Lagrangian.  This matching of the central
charges was  confirmed as that is the case in
Ref.~\cite{Nam:2010dd}. It seems that this correspondence allows one
to obtain conserved charges for new type black holes through the
formulae for the BTZ case Eq.~(\ref{AsymAdS}).  However, mass of new
type black holes obtained as conserved charge in~\cite{Oliva:2009ip}
is not consistent with this argument. We will resolve this issue by
reexamining mass of new type black holes as conserved charge by
comparing the procedure in the warped $AdS$ black hole case.

The following sections are organized as follows.  In section 2, we
briefly review the so-called ADT currents, potential and
charges~\cite{Abbott:82}\cite{Deser:2002}\cite{Deser:2003}, and
present their explicit expressions for NMG.  We give  the generic
expression of the ADT potential  for scalar curvature square and
Ricci tensor square, which is applicable in any dimension. We obtain
the vector expression for ADT charges, in $SL(2,\R)$ reduction
formalism, of (warped) $AdS$ black holes and its relation with {\it
super angular momentum} in section 3. In this section, we present
mass and angular momentum of black holes as conserved charges. This
confirms the suggested expression for these charges and verifies the
first law of black hole thermodynamics.  In section 4, we comment on
conserved charges for new type black holes.  
In the final section, we summarize our results with some comments on our approach and on the open issues.    Some calculational
details are given in the Appendices.


\section{Conserved Charges in  New Massive Gravity}
Usually, it is not so straightforward to define conserved charges in
a theory with general covariance, since the concept of conservation
requires the choice of time coordinate but  general covariance
denies the preferred time coordinate. However, for asymptotically
fixed spacetime, it seems to be manageable to define time coordinate
in a canonical way. For instance, asymptotically flat spacetime
allows so-called ADM formalism. In the case of asymptotically $AdS$
spacetime, one needs to devise another approach. In the following
subsection, we will review the approach pioneered by Abbott, Deser
and Tekin (ADT)~\cite{Abbott:82}\cite{Deser:2002}\cite{Deser:2003},
and successfully applied to the TMG
case~\cite{Clement:2007}\cite{Deser:2003cq}. In the next subsection,
we present our results of ADT charges in the NMG case.

\subsection{Review: ADT formalism}
One way to define conserved charges in gravity theory is to
construct  covariantly conserved currents and to integrate these
currents. In this subsection, we review the so-called ADT currents,
ADT charges  and their construction in terms of the linearized
Bianchi identity and a Killing vector.  This process with minimum
assumption in TMG is given in~\cite{Clement:2007}, which is adopted
and presented in our convention.

An arbitrary metric, $g_{\mu\nu}$ can be expanded around the
background metric $\bar{g}_{\mu\nu}$ as
\[ g_{\mu\nu} = \bar{g}_{\mu\nu} + h_{\mu\nu}\,, \]
where the  background metric will be taken as the solution of
equations of motion (EOM)
 \beq \CE_{\mu\nu}(\bar{g}) =0\,. \label{EOM}
 \eeq
In the following, we will call the left hand side expression of the
above equation, $\CE_{\mu\nu}$, as the EOM expression. Note that the
EOM expression need not vanish for a generic metric $g_{\mu\nu}$.

For any kind of gravitational theory without matter, the general
covariance of gravity leads to the following differential Bianchi
identity for an arbitrary metric $g_{\mu\nu}$ as
 \[ \nabla_{\mu}\CE^{\mu\nu} =0\,, \]
which imposes the linearized Bianchi identity on the linearized EOM
expression, $\delta \CE^{\mu\nu}$,  through EOM for the background
metric $\bar{g}_{\mu\nu}$,  as
 \beq
  \bar{\nabla}_{\mu}\delta \CE^{\mu\nu} =0\,.
 \eeq
Note that this expression is nothing but the equation for the
covariant conservation except that the conserved quantity is a
symmetric tensor, not a vector.  {}From now  on,   the bar notation
is dropped and so the covariant derivatives and curvatures are
referred   to the background metric for our convenience.

The ADT currents are defined by the contraction of the linearized
EOM expression, $\delta \CE$ and a Killing vector $\xi$ as
 \begin{equation}
  J^{\mu} \equiv \delta \CE^{\mu\nu}\xi_{\nu} \,.
 \end{equation}
One can see that these currents are conserved covariantly by the
linearized Bianchi identity of $\delta \CE^{\mu\nu}$ and the Killing
property of $\xi_{\nu}$. Under EOM of the background metric,
antisymmetric tensor potential, $Q^{\mu\nu}$ for this current can
defined by
\beq J^{\mu} = \nabla_{\nu}Q^{\mu\nu}\,, \eeq
which guarantees current conservation by the antisymmetric property
of the potential $Q^{\mu\nu}$. Though this is a legitimate way to
obtain the potential of currents, it is not convenient one, since we
should impose EOM to define the potential.

The current, itself, may be defined without EOM although its
conservation requires EOM, and it is not so favorable to use EOM
for a generic higher curvature theory.  Therefore, it is desirable
to define the  potential, $Q$  without EOM. That is to say, the
potential needs to be defined for a generic background. This can be
achieved by defining the antisymmetric tensor potential,
$Q^{\mu\nu}$ for the current as~\cite{Clement:2007} \beq J^{\mu} = \delta
\CE^{\mu\nu}\xi_{\nu}   \equiv
-\CE^{\mu\alpha}h_{\alpha\nu}\xi^{\nu} + \half
\xi^{\mu}\CE^{\alpha\beta}h_{\alpha\beta}-\half
h\CE^{\mu}_{\nu}\xi^{\nu}+\nabla_{\nu}Q^{\mu\nu}\,, \eeq
which reduces the previous definition of the potential when EOM is
imposed.

Noting
 \[   \delta \CE^{\mu\nu}\xi_{\nu}
 = -h^{\mu\alpha}\CE_{\alpha\nu}\xi^{\nu}-\CE^{\mu\alpha}h_{\alpha\nu}\xi^{\nu}
 +g^{\mu\alpha}\delta \CE_{\alpha\nu}\xi^{\nu}\,,
 \]
one can see that
\beq
 g^{\mu\alpha}\delta \CE_{\alpha\nu}\xi^{\nu} = h^{\mu\alpha}\CE_{\alpha\nu}\xi^{\nu}+\half \xi^{\mu}\CE^{\alpha\beta}h_{\alpha\beta}-\half h\CE^{\mu}_{\nu}\xi^{\nu}+\nabla_{\nu}Q^{\mu\nu}\,. \label{ADTpot}
\eeq
This is the form that we will utilize to obtain the ADT potential
and subsequently ADT charges.

In terms of this potential for the current, ADT charges for a
Killing vector $\xi$ is given by
\beq Q(\xi) = \frac{1}{2\kappa^2}\int d\Sigma_{\mu\nu}Q^{\mu\nu}(\xi)\,, \label{ConCharges}\eeq
where  $2\kappa^2= 16\pi G$ is the Newton's constant.

Using the ADT potential and denoting time, radial and angle
coordinates as $(t,r,\phi)$, one obtains mass and angular momentum
in the three dimensional case of our interest as
\beq M = \frac{1}{4G}\sqrt{-\det g}\,   Q^{rt}( \xi_T)   \bigg|_{ r\rightarrow\infty}\,, \qquad
J  =  \frac{1}{4G}\sqrt{-\det g}\,   Q^{rt} (\xi_R)   \bigg|_{ r\rightarrow\infty}\,,     \label{MassAng}\eeq
where  $\xi_T $ and $\xi_R$ denote the time translational and
rotational Killing vector, respectively. Note that the normalization
of Killing vectors has effects on the overall scale of conserved
charges. So, we will adopt the convention used in~\cite{Nam:2010dd}
such that in the case of asymptotic $AdS$ space of the radius $L$
normalized at the spacelike infinity as $\xi^2_T\rightarrow -r^2$
and $\xi^2_R \rightarrow L^2r^2$  with  $g_{tt}\rightarrow -L^2r^2$
and $g_{\phi\phi} \rightarrow L^2r^2$,     Killing vectors  are
taken as
\[
     \xi_T = \frac{1}{L}\frac{\p}{\p t}\,, \qquad \xi_R = \frac{\p}{\p \phi}\,,
\]
which is the convention we will use in the following.

We will illustrate the above procedure for the Einstein-Hilbert term
with cosmological constant in the Lagrangian, $\CL = R +
\frac{2}{l^2}$. The metric variation leads to EOM expression as
\[ \CE_{\mu\nu} = R_{\mu\nu} - \half g_{\mu\nu}\Big(R + \frac{2}{l^2}\Big)\,. \]
To obtain the ADT potential for ADT currents, we compute the
linearized EOM expression with the contraction of a Killing vector
$\xi$ in Eq.~(\ref{ADTpot})  as
\[
g^{\mu\rho}\xi^{\nu}\delta  \CE^{R}_{\rho\nu}
  = -\half h^{\mu}_{\nu}\xi^{\nu}\Big(R+ \frac{2}{l^2}\Big)
  +  g^{\mu\rho}\xi^{\nu} \Big(\delta  R_{\rho\nu}
  - \half g_{\mu\nu}\delta R\Big) \,.
\]
For our convenience, let us introduce
\beq
 \CI_{\mu\nu} \equiv \delta R_{\mu\nu} -\half g_{\mu\nu}\delta R\,,
\eeq
which satisfies
\bea \CI^{\mu}_{~\nu}\xi^{\nu}  &=&
 g^{\mu\alpha}(\delta R_{\alpha\beta}-\half g_{\alpha\beta}\delta R)\xi^{\beta}  \nn \\
 &=& \half \xi^{\nu}\Big(\nabla^{\alpha}\nabla^{\mu}h_{\nu\alpha}
 + \nabla^{\alpha}\nabla_{\nu} h^{\mu}_{\alpha}-\nabla^2h^{\mu}_{\nu}-\nabla^{\mu}\nabla_{\nu}h\Big)
 - \half \xi^{\mu}\Big(\nabla^{\alpha}\nabla^{\beta}h_{\alpha\beta}-\nabla^2h
 -h_{\alpha\beta}R^{\alpha\beta} \Big)\,. \nn\eea
After combining  terms to form antisymmetric total derivatives
through Leibniz's rule, one obtains
\bea \CI^{\mu}_{~\nu}\xi^{\nu}  = \nabla_{\nu}Q^{\mu\nu}_{R}
   + \half \xi^{\mu}h_{\alpha\beta}R^{\alpha\beta} + \half h \nabla^2\xi^{\mu}
   + \half \xi^{\nu} [\nabla^{\alpha},\nabla_{\nu}] h^{\mu}_{\alpha}
   + \half h^{\nu\alpha}\nabla_{\nu}\nabla^{\mu}\xi_{\alpha}
   - \half h^{\mu\alpha}\nabla^2\xi_{\alpha}\,, \eea
where
\beq
 Q^{\mu\nu}_{R} \equiv \xi_{\alpha}\nabla^{[\mu}h^{\nu]\alpha}
 - \xi^{[\mu}\nabla_{\alpha}h^{\nu]\alpha} -h^{\alpha[\mu}\nabla_{\alpha}\xi^{\nu]}
 + \xi^{[\mu}\nabla^{\nu]}h+\half h \nabla^{[\mu}\xi^{\nu]}\,. \label{EHADT}
\eeq
Using the properties of Killing vectors and the definition of
Riemann tensors, one obtains
\beq
\CI^{\mu}_{~\nu}\xi^{\nu}  =  \nabla_{\nu}Q^{\mu\nu}_{R}
    + h^{\mu\alpha}R_{\alpha\nu}\xi^{\nu} + \half \xi^{\mu}R^{\alpha\beta}h_{\alpha\beta}
    - \half R^{\mu}_{\nu}\xi^{\nu} h \,. \label{Id}
\eeq
As a result, one can see that
\[ g^{\mu\rho}\xi^{\nu}\delta  \CE^{R}_{\rho\nu}
    =  h^{\mu\alpha}\CE_{\alpha\nu}^{R} \xi^{\nu}
    +\half \xi^{\mu}\CE^{\alpha\beta}_{R}  h_{\alpha\beta}
    -\half h\CE^{ \mu}_{R\, \nu}\xi^{ \nu}+\nabla_{\nu}Q^{\mu\nu}_{R}\,,
\]
and that the ADT potential for Einstein-Hilbert terms  is given by
the above $Q^{\mu\nu}_R$.

\subsection{ADT Potential in NMG}
%
Now, let us consider ADT potentials in NMG. We will consider the
simplest version of NMG of which action is given by\footnote{We have
introduced  $\eta$ for  the various sign choice of  terms in the
action.}\cite{Bergshoeff:2009}
\begin{equation}\label{NMG}
S =\frac{\eta}{2\kappa^2}\int d^3x\sqrt{-g}\bigg[ \sigma R +
\frac{2}{l^2} + \frac{1}{m^2}K  \bigg]\,,
\end{equation}
where $\eta$ and $\sigma$ take $1$ or $-1$, and $K$ is defined by
\beq
 K = R_{\mu\nu}R^{\mu\nu} -\frac{3}{8}R^2\,.
\eeq
 Our convention is such that $m^2$ is always positive but the cosmological constant $l^2$ has no such restriction.   The EOM of NMG is given by
\begin{equation}
\CE_{\mu\nu} =\eta\Big[
 \sigma G_{\mu\nu} - \frac{1}{l^2}g_{\mu\nu} + \frac{1}{2m^2}K_{\mu\nu}\Big]
        =0\,,
\end{equation}
 where
\begin{equation}
 K_{\mu\nu} = g_{\mu\nu}\Big(3R_{\alpha\beta}R^{\alpha\beta}-\frac{13}{8}R^2\Big)
                + \frac{9}{2}RR_{\mu\nu} -8R_{\mu\alpha}R^{\alpha}_{\nu}
                + \half\Big(4\nabla^2R_{\mu\nu}-\nabla_{\mu}\nabla_{\nu}R
                -g_{\mu\nu}\nabla^2R\Big)\,.
\end{equation}

 Since the ADT potential is additive, it is sufficient to present the ADT potential for  $K$-term. The ADT potential for the $K$-term may also be obtained by adding the contribution from $R^2$ term and $R_2 \equiv R_{\mu\nu}R^{\mu\nu}$ term. In the following we present the ADT potential for $R^2$  and $R_2$, respectively.  At this stage our results are independent of dimension.
Now, let us consider each contribution separately.
For the $R^2$ term, the EOM expression is given by
\beq
 \CE_{\mu\nu}^{R^2}  = 2\Big[RR_{\mu\nu}+g_{\mu\nu}\nabla^2R-\nabla_{\mu}\nabla_{\nu}R\Big]-\half g_{\mu\nu} R^2\,.
\eeq
It is straightforward to obtain its variation as \bea
\half \delta   \CE_{\mu\nu}^{R^2} 
&=&h_{\mu\nu}\Big(\nabla^2R -\frac{1}{4}  R^2\Big) +  \delta RR_{\mu\nu}+g_{\mu\nu}\nabla^2\delta R -\nabla_{\mu}\nabla_{\nu}\delta R +R\CI_{\mu\nu}\nn \\
&& +g_{\mu\nu}\Big[- h_{\alpha\beta}\nabla^{\alpha}\nabla^{\beta}R -g^{\alpha\beta}\delta \Gamma_{\alpha\beta}^{\nu}\nabla_{\nu}R\Big] + \delta \Gamma_{\mu\nu}^{\alpha}\nabla_{\alpha}R\,. \nn
\eea
Noting that
\beq
\xi^{\mu}\nabla^2\delta R -\xi^{\nu}\nabla^{\mu}\nabla_{\nu}\delta R = \nabla_{\nu}\Big[2\xi^{[\mu}\nabla^{\nu]}\delta R + \delta R\nabla^{[\mu}\xi^{\nu]}\Big] - \delta R R^{\mu}_{\nu}\xi^{\nu}\,, \nn
\eeq
one can see that
\begin{eqnarray}
 \half g^{\mu\alpha}\delta \CE^{R^2}_{\alpha\beta}\xi^{\beta}  
 &=& \nabla_{\nu}\bigg[2\xi^{[\mu}\nabla^{\nu]}\delta R + \delta R\nabla^{[\mu}\xi^{\nu]}\bigg]  + h^{\mu}_{\nu}\xi^{\nu}\Big(\nabla^2R-\frac{1}{4}R^2\Big) +R\CI^{\mu}_{~\nu}\xi^{\nu}   \nn \\
 &&  -\xi^{\mu}\Big[h_{\alpha\beta}\nabla^{\alpha}\nabla^{\beta}R+g^{\alpha\beta}\delta \Gamma^{\nu}_{\alpha\beta}\nabla_{\nu}R\Big] + g^{\mu\rho}\xi^{\nu}\delta \Gamma^{\alpha}_{\rho\nu}\nabla_{\alpha}R\,,
\end{eqnarray}
Using the following formula
\bea
&&-\xi^{\mu}\Big[h_{\alpha\beta}\nabla^{\alpha}\nabla^{\beta}R+g^{\alpha\beta}\delta \Gamma^{\nu}_{\alpha\beta}\nabla_{\nu}R\Big] + g^{\mu\rho}\xi^{\nu}\delta \Gamma^{\alpha}_{\rho\nu}\nabla_{\alpha}R \nn \\
&&= -\nabla_{\nu}\Big[\xi^{[\mu}h^{\nu]\alpha}\nabla_{\alpha}R\Big] +Q^{\mu\nu}_R\nabla_{\nu}R -h^{\mu\alpha}\xi^{\nu}\nabla_{\alpha}\nabla_{\nu}R + \half h\xi^{\nu}\nabla^{\mu}\nabla_{\nu}R -\half \xi^{\mu}h^{\alpha\beta}\nabla_{\alpha}\nabla_{\beta}R\,, \nn \eea
and recalling the identity~(\ref{Id}) with Leibniz's rule,  one can see that the ADT  potential for the contribution of $R^2$ term is given by
\begin{eqnarray}
 Q^{\mu\nu}_{R^2} &=& 2R Q^{\mu\nu}_{R} + 4\xi^{[\mu}\nabla^{\nu]}\delta R+2\delta R \nabla^{[\mu}\xi^{\nu]} -2\xi^{[\mu}h^{\nu]\alpha}\nabla_{\alpha}R\,. \label{RsqADT}
\end{eqnarray}
where $Q^{\mu\nu}_R$ is defined in Eq.~(\ref{EHADT}) and $\delta R$ means that
\[ \delta R \equiv
-R^{\alpha\beta}h_{\alpha\beta}+\nabla^{\alpha}\nabla^{\beta}h_{\alpha\beta}-\nabla^2h\,. \]

Now, let us turn to the  contribution of $R_2\equiv R_{\mu\nu}R^{\mu\nu}$ term to the ADT potential  in $K$-term. The EOM expression of $R_2$ term is given by
\beq \CE_{\mu\nu}^{R_2}  = -2R_{\mu\alpha\beta\nu}R^{\alpha\beta}+\nabla^2R_{\mu\nu}-\nabla_{\mu}\nabla_{\nu}R + \half g_{\mu\nu}\Big(\nabla^2R-R_{\alpha\beta}R^{\alpha\beta}\Big)\,, \eeq
and its variation is
\bea \delta  \CE_{\mu\nu}^{R_2}  
&=& \half h_{\mu\nu}\Big(\nabla^2R-R_{\alpha\beta}R^{\alpha\beta}\Big)   -2(\delta R_{\mu\alpha\beta\nu}R^{\alpha\beta} +R_{\mu\alpha\beta\nu} \delta R^{\alpha\beta}) - \half g_{\mu\nu} \delta (R_{\alpha\beta}R^{\alpha\beta})   \\
&& + g_{\mu\nu}\nabla^2\delta R -\nabla_{\mu}\nabla_{\nu}\delta R + \nabla^2\CI_{\mu\nu}  -\nabla^{\alpha}(h_{\alpha\beta}\nabla^{\beta}R_{\mu\nu})+\half\nabla_{\alpha}h\nabla^{\alpha}R_{\mu\nu} + F_{\mu\nu} \nn \\
&& -  2\delta \Gamma^{\beta}_{\alpha\mu}\nabla^{\alpha}R_{\beta\nu}  - 2\delta \Gamma^{\beta}_{\alpha\nu}\nabla^{\alpha}R_{\beta\mu}  + \half g_{\mu\nu}\Big[-\nabla^{\alpha}(h_{\alpha\beta}\nabla^{\beta}R) + \half \nabla_{\alpha}h\nabla^{\alpha}R\Big] + \delta \Gamma^{\alpha}_{\mu\nu}\nabla_{\alpha}R\,,  \nn  \eea
where
\beq  F_{\mu\nu}  \equiv
-\nabla^{\alpha}\delta \Gamma^{\beta}_{\alpha\mu}R_{\beta\nu} - \nabla^{\alpha}\delta \Gamma^{\beta}_{\alpha\nu}R_{\beta\mu} \,.  \label{Fterm}
\eeq

To obtain ADT potential according to~(\ref{ADTpot}) in the squared Ricci curvature  case, let us write the contraction of linearized EOM expression with a Killing vector  as
\beq
g^{\mu\rho} \delta \CE_{\rho\nu}\xi^{\nu}  =  \half h^{\mu}_{\nu}\xi^{\nu}\Big(\nabla^2R-R_{\alpha\beta}R^{\alpha\beta}\Big) +   E^{\mu}_{1} + E^{\mu}_{2} + E^{\mu}_{3} + E^{\mu}_{4} + E^{\mu}_{5}\,, \eeq
where
\bea
 E^{\mu}_{1} &\equiv&
-2g^{\mu\rho}\xi^{\nu}\delta (R_{\rho\alpha\beta\nu} R^{\alpha\beta}) - \half \xi^{\mu}\delta (R_{\alpha\beta}R^{\alpha\beta})     - 2\xi^{\nu}\delta \Gamma^{\beta}_{\alpha\nu}\nabla^{\alpha}R_{\beta}^{\mu}  \,, \nn \\
E^{\mu}_{2} &\equiv & \xi^{\mu}\nabla^2\delta R -\xi^{\nu}\nabla_{\nu}\nabla^{\mu}\delta R + \xi^{\nu}\nabla^2\CI^{\mu}_{\nu} -g^{\mu\rho}\xi^{\nu}R_{\nu\beta}\nabla^{\alpha}\delta \Gamma^{\beta}_{\alpha\rho} - \xi^{\nu}R_{\beta}^{\mu}\nabla^{\alpha}\delta \Gamma^{\beta}_{\alpha\nu}    \,, \nn \\
E^{\mu}_{3}
& \equiv & -  2g^{\mu\rho}\xi^{\nu}\delta \Gamma^{\beta}_{\alpha\rho}\nabla^{\alpha}R_{\beta\nu}  - \xi^{\nu}\nabla_{\alpha}(h^{\alpha\beta}\nabla_{\beta}R^{\mu}_{\nu})\,, \nn \\
E^{\mu}_{4}&\equiv&  -\half \xi^{\mu}\nabla_{\alpha}(h^{\alpha\beta}\nabla_{\beta}R) + g^{\mu\rho} \xi^{\nu} \delta \Gamma_{\rho\nu}^{\alpha}\nabla_{\alpha}R\,,
 \nn \\
 E^{\mu}_{5}  &\equiv& \half \xi^{\nu} \nabla_{\alpha}h\nabla^{\alpha}R^{\mu}_{\nu}  + \frac{1}{4}\xi^{\mu} \nabla_{\alpha}h\nabla^{\alpha}R\,. \nn
\eea
Collecting various contributions from each $E^{\mu}_{a}$ term which
can be obtained by using formulae in the Appendix A, one can obtain the contribution of the
squared Ricci curvature $R_{2} = R_{\mu\nu}R^{\mu\nu}$
to the ADT potential as
\bea Q^{\mu\nu}_{R_2}
&=&
2\xi^{\alpha}R^{\beta[\mu}\nabla_{\beta}h^{\nu]}_{\alpha}-2\xi^{\alpha}R^{\beta[\mu}\nabla^{\nu]}h_{\alpha\beta}+2R^{\alpha\beta}\xi^{[\mu}\nabla^{\nu]}h_{\alpha\beta}-2R^{\alpha\beta}\xi^{[\mu}\nabla_{\alpha}h^{\nu]}_{\beta} +2h^{\alpha\beta}\nabla_{\alpha}(\xi^{[\mu}R^{\nu]}_{\beta}) \nn \\
&&  +  R^{\mu\nu}_{~~\, \alpha\beta} \xi^{\alpha} \nabla_{\gamma}h^{\gamma\beta}-2\xi^{\gamma}R_{\gamma}^{~\alpha\beta[\mu}\nabla_{\alpha}h^{\nu]}_{\beta} + 2\xi^{\gamma}R_{\gamma}^{~\alpha\beta[\mu}\nabla^{\nu]}h_{\alpha\beta}
  \nn \\
&& + \nabla^2Q^{\mu\nu}_{R} + R^{\mu\nu}_{~~\alpha\beta}Q^{\alpha\beta}_{R} +2Q^{\alpha[\mu}_RR^{\nu]}_{\alpha}+2\xi^{[\mu}\nabla^{\nu]}\delta R -2\xi^{\alpha}R_{\alpha}^{[\mu}\nabla_{\beta}h^{\nu]\beta}     -2\nabla^{\alpha}\xi^{\beta} \nabla_{\alpha}\nabla^{[\mu}h^{\nu]}_{\beta} \nn \\
&& + 2\xi^{\beta}h^{\alpha[\mu}\nabla_{\alpha}R^{\nu]}_{\beta}  -\xi^{[\mu}h^{\nu]\alpha}\nabla_{\alpha}R
\nn \\
&& + 2R_{\alpha\beta}h^{\alpha[\mu}\nabla^{\nu]}\xi^{\beta} -2 h_{\alpha\beta}R_{\gamma}^{~\alpha\beta[\mu}\nabla^{\nu]}\xi^{\gamma} \nn \\
&& -R^{\mu\nu}_{~~\alpha\beta}\xi^{[\alpha}\nabla^{\beta]}h + \xi^{\alpha}R_{\alpha}^{[\mu}\nabla^{\nu]}h + \xi^{[\mu}R^{\nu]\alpha}\nabla_{\alpha} h -2hR_{\alpha}^{[\mu}\nabla^{\nu]}\xi^{\alpha}\,.
\eea
Using three dimensional identity,
 $R^{\mu\nu}_{~~\,\alpha\beta} =2(\delta^{[\mu}_{\alpha}R^{\nu]}_{\beta}   -  \delta^{[\mu}_{\beta}R^{\nu]}_{\alpha}) -  R\delta^{[\mu}_{\alpha}\delta^{\nu]}_{\beta}
$, one may rewrite $Q^{\mu\nu}_{R_2}$ as
\bea Q^{\mu\nu}_{R_2} &=& \nabla^2Q^{\mu\nu}_R +\half Q^{\mu\nu}_{R^2} -2Q^{\alpha[\mu}_RR^{\nu]}_{\alpha} -2\nabla^{\alpha}\xi^{\beta}\nabla_{\alpha}\nabla^{[\mu}h^{\nu]}_{\beta}   -4\xi^{\alpha}R_{\alpha\beta}\nabla^{[\mu}h^{\nu]\beta}
   -Rh_{\alpha}^{[\mu}\nabla^{\nu]}\xi^{\alpha}  \nn  \\
&&  + 2 \xi^{[\mu}R^{\nu]}_{\alpha}\nabla_{\beta}h^{\alpha\beta}     + 2\xi_{\alpha}R^{\alpha[\mu}\nabla_{\beta}h^{\nu]\beta} + 2\xi^{\alpha}h^{\beta[\mu}\nabla_{\beta}R^{\nu]}_{\alpha} +2h^{\alpha\beta}\xi^{[\mu}\nabla_{\alpha}R^{\nu]}_{\beta}   \nn \\
&& -(\delta R +2R^{\alpha\beta}h_{\alpha\beta})\nabla^{[\mu}\xi^{\nu]}    -3\xi^{\alpha}R_{\alpha}^{[\mu}\nabla^{\nu]}h -\xi^{[\mu}R^{\nu]\alpha}\nabla_{\alpha}h\,.   \label{RiccisqADT}\eea
One of our main results is the explicit expression for the
ADT potential for the $K$-term in NMG, which is given by the sum of two expressions in Eq.~(\ref{RsqADT}) and Eq.~(\ref{RiccisqADT}) as
\beq
 Q^{\mu\nu}_{K}  =  Q^{\mu\nu}_{R_2}-\frac{3}{8}Q^{\mu\nu}_{R^2}  \,.
\eeq

In the following sections, we give the ADT potential
$Q^{\mu\nu}_{R}$ and $Q^{\mu\nu}_K$ for each Killing vector, $\xi_T$
and $\xi_R$ on various black holes.  The mass and angular momentum
obtained by these potentials confirm suggested expression for those
quantities.

\section{Mass and Angular Momentum as Conserved Charges}

In this section we will relate  ADT charges or potential to {\it
super angular momentum} by Clement. Before presenting our results,
we review for completeness  briefly  $SL(2,\R)$ reduced action
method which is useful particularly for the three dimensional
gravity~\cite{Clement:2009gq}\cite{Clement:1994}\cite{Clement:2008}
After that, we present the so-called correction term to mass in {\it
super angular momentum} formalism by using the explicit expression
of ADT potential.

\subsection{$SL(2,\R)$ Reduction Method}
Let us take the three dimensional metric  ansatz as
\beq
 ds^2 = \lambda_{ab}(\rho)dx^{a}dx^{b} + \frac{d\rho^2}{\zeta^2 U^2(\rho)}\,, \qquad x^a=(t, \phi). \label{SLmetric}
\eeq
where all variables are functions only of the radial coordinate
$\rho$.   By the reparametrization invariance with respect to $\rho$
coordinate, function $U$ may be chosen such that the condition $\det
\lambda = -U^2$ is satisfied. This choice implies that $\sqrt{-g} =
1/\zeta$.    This metric ansatz  reduces the given generally
covariant Lagrangian to the $SL(2,\R)\simeq SO(1,2)$ invariant one.
This means that the relevant variables becomes the $SO(1,2)$ three
dimensional vectors and differential geometric calculation reduces
to the vector calculus.  In particular, Einstein equations  reduces
to the equations obtained by the variation with respect to  the two
dimensional metric $\lambda$ and the one with respect to $\zeta$.
These equations are called EOM and Hamiltonian constraint,
respectively in this $SL(2,\R)$ reduction formalism. Using the
vector expression for EOM, one may integrate EOM to obtain some
conserved quantities, which is the analog of angular momentum in
mechanical problem: This quantity  is named as {\it super angular
momentum}~\cite{Clement:1994gr}.

Explicitly, one may parameterize generic symmetric two by two matrix
$\lambda$ as
\[ \lambda_{ab} = \left(\ba{cc} X^0+X^1 & X^2 \\ X^2 & X^0 -X^1 \ea\right)\,,
\]
and associate a  $SO(1,2)$  three dimensional vector ${\bf X}=(X^0,X^1,X^2)$ with this matrix. Conversely, the associated matrix with a vector ${\bf X}$ is denoted by $\langle{\bf X}\rangle$ and given explicitly by
\[ \langle{\bf X}\rangle=\bftau\cdot {\bf X} = \left(\ba{cc} -X^2 & -X^{0}+X^{1} \\ X^{0}+X^{1} & X^{2} \ea\right)\,, \]
where
\[
\tau^0 = \left( \ba{rr}  0 &  -1 \\ 1 & 0 \ea\right)\,, \qquad \tau^1 = \left( \ba{rr}  0 &  1 \\ 1 & 0 \ea\right)\,, \qquad \tau^2 = \left( \ba{rr}  -1 &  0 \\ 0 & 1 \ea\right)\,. \]
Note that ${\bf X}^2 = - \det \lambda = U^2$.
 Then,  the above matrix $\lambda$ can be written as $\lambda = \tau^0\langle{\bf X}\rangle$ and its inverse  is given by
$\lambda^{-1} = -U^{-2}\langle{\bf X}\rangle\tau^0$.   In terms of $\lambda$ and $U$, the connections of the above metric ansatz is given by
\[
 \Gamma^{\rho}_{ab} = -\frac{\zeta^2}{2} U^2 \lambda'_{ab}\,, \qquad \Gamma^{b}_{\rho a}=\half (\lambda^{-1}\lambda')^{b}_{a}   \,, \qquad \Gamma^{\rho}_{\rho\rho} = -\frac{U'}{U}\,, \qquad \qquad (~ ' \equiv \frac{d}{d \rho})
\]
which are subsequently written in terms of ${\bf X}$ and its
derivatives. All other relevant differential geometric quantities
such as the deviation metric $h$ and  curvatures can be written in
terms of vector ${\bf X}$ and its derivatives.   For this purpose,
it is useful  to define the inner and cross product of $SO(1,2)$
vectors  in the standard way as
\[
{\bf A}\cdot {\bf B} = \eta_{ij}A^{i} B^{j}\,, \qquad ({\bf A}\times {\bf B})^i = \eta^{im}\epsilon_{mjk}A^{j} B^{k}\,, \qquad \qquad (\epsilon_{012} =1)  \]
and note that the product of two  matrices dual to vectors ${\bf A}$ and ${\bf B}$ is given by
\[
 \langle {\bf A}\rangle  \langle {\bf B} \rangle = ({\bf A}\cdot {\bf B})\, {\bf 1} + \langle{\bf A}\times {\bf B}\rangle\,.
\]
For instance, one can see that\footnote{From now on, we will drop the identity matrix notation, ${\bf 1}$ just for notational convenience.},
\bea
 h^{a}_{b} &=& (\lambda^{-1}\delta \lambda)^{a}_{b}=  \frac{\delta U}{U} + \frac{1}{U^2}\langle {\bf \Sigma} \rangle \,, \qquad h^{\rho}_{\rho} = -2\frac{\delta U}{U}\,, \nn \\
  \lambda^{-1}\lambda' &=& \frac{U'}{U} + \frac{1}{U^2} \langle {\bf L} \rangle  \,,  \qquad \qquad \qquad \lambda' \lambda^{-1}= \tau^0\Big[-\frac{U'}{U}  +  \frac{1}{U^2}\langle {\bf L} \rangle  \Big]  \tau^0 \,, \nn \eea
where ${\bf L}$ and ${\bf \Sigma}$ are defined by
\[  {\bf L} \equiv {\bf X} \times {\bf X}' \,, \qquad {\bf \Sigma} \equiv {\bf X}\times \delta {\bf X}\,. \]
Some other vector representation of geometric quantities and useful vector identities are given in the Appendix B.

For the given metric, the Killing vector is taken as $\xi^{\mu} \equiv  (k^a,0)$, which include $\xi_T$ and $\xi_R$ as the special cases. For this Killing vector,  one obtains
\[
 \nabla_{\rho}\xi_{a} =-\nabla_{a}\xi_{\rho}= \half (k\lambda')_{a} \,, \qquad \nabla_{a}\xi_{b}=0  \,.
\]
Now, one can see that the ADT potentials for Einstein-Hilbert term,
of which specific component leads to mass and angular momentum, are
given by
\bea Q^{ab}_R &=&0\,, \nn \\
Q^{\rho a}_R &=& \frac{\zeta^2}{2} U^2 \bigg[k\lambda\Big(\nabla_{\mu}h^{\mu}_{\rho}  +  (\lambda^{-1}\delta\lambda)' + 2\frac{\delta U}{U}\lambda^{-1}\lambda'\Big)\lambda^{-1}\bigg]^{a}  \nn \\
&=& \frac{\zeta^2}{2}\bigg[ - k\, ({\bf X}\cdot \delta {\bf X}') +  k\tau^0 \langle \delta{\bf L} \rangle    \tau^0 \bigg]^{a} \,,  \eea
where we used various formulae given in the Appendix B.
Note also that
\beq \delta R \equiv \nabla^{\mu}\nabla^{\nu}h_{\mu\nu}-\nabla^2 h - R^{\mu\nu}h_{\mu\nu} =-\zeta^2\Big[3{\bf X}'\cdot \delta {\bf X}' + 2{\bf X}\cdot \delta {\bf X}'' + 2\delta {\bf X}\cdot {\bf X}''\Big]\,. \eeq

Since the vector representation of the ADT potential for $K$ term is
somewhat involved, calculational details are relegated to the
Appendix C. Collecting all the terms in Appendix C and using vector
identities in the Appendix B, one can obtain the vector
representation of the ADT potential for $K$ term and then see that
\bea
\eta \bigg[\sigma Q_R^{\rho t} + \frac{1}{m^2}Q_{K}^{\rho
t}\bigg]_{\xi_T}  &=&   \frac{1}{L} \bigg[- \frac{\zeta^2}{2}~
\delta  J^{2} + \Delta_{Cor}\bigg]  \,,  \label{MassADT} \\
\eta \bigg[\sigma Q_R^{\rho t} + \frac{1}{m^2}Q_{K}^{\rho
t}\bigg]_{\xi_R} &=&  \frac{\zeta^2}{2}~ \delta ( J^{0}-J^{1})  \,,
\label{AngADT}
\eea
where  ${\bf J}= (J^0, J^1, J^2)$ is given by
\beq\label{SupAng}
 {\bf J} = \eta\Bigg(\sigma\, {\bf L}  + \frac{\zeta^2}{m^2}\bigg[2{\bf L}\times {\bf L}' + {\bf X}^2\,   {\bf L}''  +  \frac{1}{8}\Big({\bf X}'^2 -4 {\bf X}\cdot {\bf X}''\Big){\bf L}\bigg]  \Bigg)\,.   \label{SuperAng}
\eeq
The above ${\bf J}$ vector is the so-called {\it super angular
momentum} in NMG, and $\Delta_{Cor}$ is the so-called {\it
correction term} to mass. The correction term to mass is composed of
two parts, one of which comes from Einstein-Hilbert term and the
other from $K$-term as
\beq \Delta_{Cor}  = \Delta_{R} + \Delta_{K}\,. \eeq
Each term is given by
\bea \Delta_{R} &=& \eta\sigma\frac{\zeta^2}{2}\Big[ -({\bf X}\cdot
\delta {\bf X}')\Big]\,, \label{Correction1}
\\
\Delta_{K} &=& \eta\frac{\zeta^4}{m^2}\bigg[-U^2 ({\bf X}''\cdot \delta {\bf X}')  +  \frac{U^2}{2}\Big[ (U\delta U)''' - ({\bf X}\cdot \delta {\bf X}')'' - \half ({\bf X}'\cdot \delta {\bf X}')' \Big]  \label{Correction2}  \\
 &&   ~~~~~~~    -\frac{UU'}{4}\Big[(U\delta U)''     -  \frac{5}{2}({\bf X}'\cdot \delta {\bf X}') \Big]    + \Big[ {\bf X}'^2-(UU')'\Big] (U\delta U)'     + UU'\, ( {\bf X}''\cdot \delta {\bf X})   \nn \\
 && ~~~~~~~   + \Big[ \frac{5}{4}(UU')' - \frac{21}{16}{\bf X}'^2\Big] ({\bf X}\cdot \delta {\bf X}')   + \Big[ -\frac{1}{2}(UU')'' + \frac{9}{4}({\bf X}'\cdot {\bf X}'' ) \Big]~ U\delta U \bigg]\,.  \nn \eea
Finally,  mass and angular momentum in NMG can be obtained by
\bea M & =&  \frac{1}{4G}\sqrt{-\det g}\, \eta\Big[ \sigma Q^{rt}_{R}  +  \frac{1}{m^2}Q^{rt}_{K}\Big]_{\xi_T,\, r\rightarrow\infty}\,,  \\
J  &=&  \frac{1}{4G}\sqrt{-\det g}\, \eta\Big[ \sigma Q^{rt}_{R}  +  \frac{1}{m^2}Q^{rt}_{K}\Big]_{\xi_R, \, r\rightarrow\infty}\,.   \eea

The above {\it super angular momentum}  has been obtained  by integrating the EOM of NMG once, as~\cite{Clement:2009gq}
\beq {\bf J} = {\bf L} -\frac{\zeta^2}{m^2}\bigg[{\bf X}^2({\bf X}\times {\bf X}''' - {\bf X}'\times {\bf X}'') +2({\bf X}\cdot {\bf X}')\, {\bf L}' +  \Big(\frac{1}{8}{\bf X}'^2 - \frac{5}{2}{\bf X}\cdot {\bf X}''\Big){\bf L}\bigg]\,, \eeq
which is identical with the above form of the {\it super angular momentum} under the choice\footnote{This is the sign choice in~\cite{Clement:2009gq}. {}From now on, this sign choice is assumed for the $WAdS_3$ case.} of $\eta=\sigma =-1$ as one can verify from the vector identities in the Appendix B.
%

\subsection{Mass and Angular momentum of BTZ and Warped $AdS_3$ Black Holes}

Through the $SL(2,\R)$ reduction,  black hole solutions may be
represented in terms of a $SO(1,2)$ vector ${\bf X}$. More
explicitly,  the EOM and Hamiltonian constraint in the NMG case
given as (2.14) and (2.15)  in~\cite{Clement:2009gq}  can be solved
by a suitable choice of a vector ${\bf X}$, which in turn leads  to
black hole solutions. Since BTZ and three dimensional warped $AdS$
($WAdS_3$) black holes are already obtained in this
way~\cite{Clement:2007}\cite{Clement:2003}, we give their
corresponding vectors in the following and then present their mass
and angular momentum.

Using $SO(1,2)$ vectors corresponding BTZ and $WAdS_3$ black holes,
we present  mass and angular momentum for each black holes. First,
BTZ black holes can be represented by\footnote{This is not unique
choice for two vectors, since one may perform  $SO(1,2)$
transformation to these.}
\beq {\bf X} = \rho \bfeta + \bfxi\,. \eeq
where
\bea \label{BTZ-vec}
  \bfeta = (0,-L^2,0)\,, \qquad \bfxi =
  \Big(\frac{L^2}{2}(\rho_++\rho_-),\,  \frac{L^2}{2}(\rho_++\rho_-),
   \, L^2\sqrt{\rho_+\rho_-} \Big)\,.
\eea
This vector ${\bf X}$ with   $\zeta^2=4/L^6$   leads to  BTZ metric as
\beq ds^2 =L^2\bigg[ -\frac{(\rho-\rho_+)(\rho - \rho_-)}{\rho}dt^2 + \rho\Big(d\phi + \frac{\sqrt{\rho_+\rho_-}}{\rho}dt\Big)^2 + \frac{d\rho^2}{4(\rho-\rho_+)(\rho-\rho_-)}\bigg]\,, \eeq
and {\it super angular momentum} as
\bea
  {\bf J} =  \eta\bigg[\sigma   +  \frac{\zeta^2}{8m^2}\,  \bfeta^2\bigg]  (\bfxi\times \bfeta)   = \eta \Big( \sigma +\frac{1}{2m^2L^2} \Big)  \Big[ L^4 \sqrt{\rho_+
         \rho_-} \bfepsilon_1 -  \frac{L^4}{2}(\rho_+ + \rho_-) \bfepsilon_2 \Big]\,, \eea
where $\bfepsilon_1 = (1,0,0)$ and $\bfepsilon_2 = (0,0,1)$.  One
may note that EOM and Hamiltonian constraint lead to $ L^2  = l^2/2 \cdot
(\sigma \pm \sqrt{1-1/m^2l^2}) $  in this case.

To obtain $WAdS_3$ black holes, one takes the $SO(1,2)$ vector ${\bf X}$ as
\[ {\bf X} = \rho^2\bfalpha + \rho \bfbeta + \bfgamma\,, \]
where $\bfalpha,\bfbeta,\bfgamma$ can be chosen as
\[ \bfalpha = \Big({\textstyle \half}, {\textstyle -\half},0\Big)\,, \qquad \bfbeta = \Big(\omega,-\omega,-1\Big)\,, \quad \bfgamma=2u\bfalpha + (1-\beta^2)\Big( {\textstyle \half},{\textstyle \half},-\omega\Big)\,. \]
Here, $u$ is defined as $ 2u \equiv  (1-\beta^2)\omega^2 + \frac{\beta^2\rho^2_0}{1-\beta^2}$.
These vectors form  a $SO(1,2)$ basis and
satisfy the following properties
\bea &&\bfalpha^2 =0\,, \quad \bfbeta^2 =1\,, \quad \bfgamma^2 = -\beta^2\rho^2_0 \,,  \quad \bfalpha\cdot \bfbeta =0\,, \quad \bfalpha\cdot \bfgamma = -\half(1-\beta^2)\,, \quad \bfbeta\cdot \bfgamma =0\,,  \nn \\
&& \bfalpha \times \bfbeta = -\bfalpha\,, \quad \bfbeta \times \bfgamma = \frac{2\beta^2\rho^2_0}{1-\beta^2}\bfalpha -\bfgamma\,, \quad \bfgamma \times \bfalpha =\half(1-\beta^2)\bfbeta\,. \nn  \eea

This $SO(1,2)$ vector ${\bf X}$ leads to the $WAdS_3$ black holes
metric as~\cite{Clement:1994gr}\cite{Clement:2009gq}
\begin{eqnarray}
 ds^2 = -\frac{\beta^2(\rho^2-\rho_0^2)}{Z^2}dt^2
        + \frac{d\rho^2}{\zeta^2\beta^2(\rho^2 - \rho_0^2)}
        + Z^2 \left( d\phi - \frac{\rho
        + (1-\beta^2)\omega}{Z^2}dt \right)^2  \,,
\end{eqnarray}
where $Z^2$ is defined by
\begin{eqnarray}
Z^2
          = \rho^2 + 2\omega \rho + (1-\beta^2)\omega^2
          + \frac{\beta^2 \rho_0^2}{1-\beta^2} \,,
\end{eqnarray}
and to satisfy the Hamiltonian constraint in this case,   $\zeta$
should be taken as
 $\zeta^2/m^2 = 8/(21-4\beta^2)$.
Note that for this vector ${\bf X}$ one obtains
\beq {\bf L} = \Big(\rho^2 -\frac{2\beta^2\rho^2_0}{1-\beta^2}\Big) \bfalpha + (1-\beta^2)\rho \bfbeta + \bfgamma\,,   \qquad
{\bf L}\times {\bf L}' = (1-\beta^2){\bf X}\,.
\eeq
The background metric and  the related basis  vectors
$\bfalpha_B,\bfbeta_B,\bfgamma_B$ can  also   be defined by setting
$\omega = \rho_0 =0$ in the above expression and, satisfies the
similar properties.

Before presenting the result about mass and angular momentum of BTZ
and $WAdS_3$ black holes in NMG, we need some remarks to  clarify
the super angular momentum approach focusing on the BTZ case  of
pure Einstein-Hilbert gravity with cosmological constant. The {\it
super angular momentum} of pure Einstein-Hilbert  gravity is simply
given by
\[ {\bf L}  = {\bf X} \times {\bf X}'  \,. \]
Since this is obtained by integrating EOM once in the vector
representation and can be shown to be conserved, in general it may
contain the non-linear effects.  That is to say, at the non-linear
level,   conserved charges are related to  subtraction of the
background values from the black hole ones
\[ \Delta {\bf L} \equiv {\bf L} -{\bf L}_B = \Delta {\bf X}\times {\bf X}'_B
      + {\bf X}_B\times \Delta {\bf X}' +  \Delta {\bf X}\times \Delta {\bf X}'\,. \]
Since ${\bf X}$ is already linear in some sense, the linearized
version of this subtraction in  {\it super angular momentum}, which
will be denoted as $\delta {\bf L}$, may be taken as the sum of the
first two terms in the above expression for $\Delta {\bf L}$.
Incidentally in the BTZ case,  the last term vanishes automatically
and there is no difference between linear and non-linear version.
Noting that the contribution of correction terms  to mass vanishes
in the case of BTZ black holes ,  one can see that  mass and angular
momentum of BTZ black holes are given by
\beq M = \frac{\rho_+ + \rho_-}{8G} \eta \Big(\sigma + \frac{1}{2m^2L^2}\Big)\,, \qquad
     J = \frac{L\sqrt{\rho_+\rho_-}}{4G} \eta \Big(\sigma + \frac{1}{2m^2L^2}\Big)\,, \eeq
which is consistent with the results from other methods~\cite{Nam:2010dd}

However, the situation is different in the case of $WAdS_3$ black
holes. To get the correct angular momentum   in this case, one
should consider the full non-linear effects. Therefore, one should
elevate the expression of the ADT potential  for angular momentum in
Eq.~(\ref{AngADT})  to the non-linear subtraction form. Concretely,
the angular momentum expression is read as
\beq
J = \frac{1}{4G}\sqrt{-\det g}\bigg[ \frac{\zeta^2}{2}~ \Delta ( J^{0}-J^{1}) \bigg] \,.
\eeq
It is interesting to note that this procedure has no effect on mass
computation. As we will see in the next section, it turns out that
even mass needs non-linear correction in new type black holes.

One way to understand the necessity of this non-linear completion
for angular momentum and its irrelevance for mass in the case of
$WAdS_3$ black holes is to pay attention to the contribution from
Einstein-Hilbert term and $K$-term separately in the Lagrangian.
Though each contribution to the angular momentum from the
Einstein-Hilbert term and the $K$-term is divergent when $\rho$ goes
to the infinity, the sum of them is not divergent and gives us the
next leading finite contribution, whereas this is not the case for
mass since each contribution is finite. Recalling that only the
linear metric deviation $h_{\mu\nu}$ is captured in our formalism of
the ADT potential, the cancelation of the leading contribution of
the ADT potential indicates strongly that it may be insufficient to
consider only the linear level contribution. Specifically, the
quadratic order contribution  in $h_{\mu\nu}$ would leads some
corrections to the finite result of angular momentum.   Without
resorting to another formalism,  one can obtain the correct value of
angular momentum through {\it super angular momentum},  since {\it
super angular momentum} is the complete non-linear integrated
charges.

Taking into account  the previous remark and noting
\[
 \Delta \bfalpha =0\,, \quad \Delta \bfbeta = 2\omega\bfalpha\,, \quad \Delta \bfgamma = 2\Big(u-(1-\beta^2)\omega^2\Big)\bfalpha +(1-\beta^2)\omega\bfbeta\,,
\]
one obtains
\bea \Delta {\bf J} &=&\Delta {\bf L} -\frac{\zeta^2}{m^2}\bigg[ 2(1-\beta^2)\Delta {\bf X} -2\beta^2\rho^2_0 \bfalpha + \frac{1}{8}(5-4\beta^2)\Delta {\bf L}\bigg] \nn \\
&=& \frac{16}{21-4\beta^2}\bigg[-\Big(\beta^2(1-\beta^2)\omega^2 + \frac{\beta^2\rho^2_0}{1-\beta^2} \Big)\bfalpha + \beta^2(1-\beta^2)\omega \bfbeta \bigg]  \nn \\
&=&\frac{16}{21-4\beta^2}\bigg[\beta^2\Big((1-\beta^2)\omega^2 -
\frac{\rho^2_0}{1-\beta^2} \Big)\bfalpha_B +
\beta^2(1-\beta^2)\omega \bfbeta_B \bigg]   \,,\eea
and the correction terms to mass are given by
\[
 \Delta_{R} = \frac{\zeta^2}{2}(1-\beta^2)\omega\,, \qquad \Delta_{K}
            = -\frac{\zeta^4}{16m^2}(21-20\beta^2)(1-\beta^2)\omega\,.
\]
Finally,  mass  of $WAdS_3$ black holes is given by
\beq M = \frac{4\zeta\beta^2(1-\beta^2)\omega}{G(21-4\beta^2)}\,  \label{WAdSmass}\eeq
and their angular momentum by
\beq   J = \frac{2\beta^2\zeta}{G(21-4\beta^2)}\Big[ (1-\beta^2)\omega^2 -\frac{\rho^2_0}{1-\beta^2}\Big]\,.\eeq
This result about angular momentum is already
in~\cite{Clement:2009gq}.  But, mass is obtained by assuming the
validity of the first law of black hole thermodynamics or through
$AdS/CFT$ correspondence not as conserved charge. Our explicit
computation of mass as conserved charge in Eq.~(\ref{WAdSmass})
through Eq.~(\ref{MassADT}) verify the first law of black hole
thermodynamics and checks $AdS/CFT$ correspondence in this case.
Note that the correction term to mass gives us exactly the same
value with the {\it super angular momentum} and mass becomes twice
of the value given by  {\it super angular momentum}. This  result
confirms the form given in~\cite{Clement:2009gq}\cite{Nam:2010dd}.

For comparison, let us present the linearized version of $\Delta
{\bf J}$, denoted as $\delta {\bf J}$
\[ \delta {\bf J} = \delta {\bf L} -\frac{\zeta^2}{8m^2}\bigg[\bigg\{2(1-\beta^2)(21-4\beta^2)\omega\rho -(5-4\beta^2) \Big((1-\beta^2)\omega^2 + \frac{\beta^2\rho^2_0}{1-\beta^2}\Big)\bigg\}\bfalpha_{B} +(21-20\beta^2)(1-\beta^2)\omega \bfbeta_{B}\bigg]\,.
\]
As can be seen easily, the nonlinear effects in ${\bf J}$ come from
$\Delta{\bf L}$ and $\Delta {\bf X}^2$.

\section{The Conserved Charges of New Type Black Holes}

In this section we confine ourselves to the simplest version of new
type black holes: static non-rotating ones. Though the mass of new
type black holes are already computed in the linearized ADT
formalism~\cite{Oliva:2009ip}\cite{Giribet:2009qz}, it is
inconsistent with $AdS/CFT$ correspondence as was shown
in~\cite{Nam:2010dd}. We reexamine mass of new type black holes as
conserved charges by using the formalism presented in the previous
section. As the case of $WAdS_3$ black holes, in which  the ADT
formalism needs to be supplemented by {\it super angular momentum}
to obtain correct angular momentum of black holes,  it turns out
that even mass needs  non-linear corrections in the case of new type
black holes. Therefore, we rederive new type black hole solutions in
the $SL(2,\R)$ reduction formalism, and then present our results
about mass of new type black holes.

In order to get the new type black hole solution in $SL(2, \R)$
reduction formalism, we can take the vector ansatz as a form
 \bea {\bf X} =  \rho \bfeta +  \sqrt{\rho}\bfkappa +  \bfxi\,.   \eea
For the above vectors, we need to impose the following conditions
\beq\bfkappa^2 = 0\,, \qquad \bfxi^2 = 0\,, \qquad \bfkappa \cdot \bfxi = 0\,, \qquad
(\bfeta \cdot \bfkappa)\bfxi - (\bfeta \cdot \bfxi) \bfkappa = 0\,,
   \eeq
which can be satisfied  by taking
\bea \label{Vectors}
 \bfeta = (0,-L^2,0)\,, ~~ \bfkappa = \left( -\frac{b}{2}L^2, -\frac{b}{2}L^2, 0
 \right) \,, ~~ \bfxi = \left(-\frac{c}{2}L^2, -\frac{c}{2}L^2, 0
 \right)\,.
\eea
EOM and Hamiltonian constraint are given by
\beq \left( \bfeta^2 - 8 \sigma \frac{m^2}{\zeta^2} \right) \bfkappa
    = 0\,, \qquad
    \frac{1}{32}(\bfeta^2)^2 - \sigma\frac{m^2}{2\zeta^2}
\bfeta^2 + \frac{2m^2}{\zeta^4 \l^2} = 0\,,  \label{condi2} \eeq
which lead to $\zeta^2 = 8m^2/L^4=4/L^6$ and $l^2 = 2L^2$,
respectively.   It turns out that the sign choice of $\eta=\sigma=1$
is necessary for the existence of new type black holes.
The above vector ${\bf X}$ leads to the metric of new type black
holes in the form of
\[ ds^2 =  L^2 \bigg[-(\rho + b\sqrt{\rho} + c)dt^2 +   \rho d\phi^2
        + \frac{d\rho^2}{4\rho(\rho + b\sqrt{\rho}+c)} \bigg]\,. \]
with $\zeta^2 = 4/L^6$.  By  the coordinate transformation $\rho =
r^2$, new type black hole solutions become
\bea
  ds^2 = L^2 \left[ -(r^2 + b r + c)dt^2 + \frac{dr^2}{(r^2 + b r + c)} + r^2 d\phi^2
  \right] \,,
\eea
which is the form given in~\cite{Bergshoeff:2009aq}.
To obtain mass of new type black holes, we compute the {\it super
angular momentum} and correction terms to mass. By inserting the
above three vectors $\bfeta$, $\bfkappa$ and $\bfxi$ into the
formula (\ref{SupAng}),we obtain {\it super angular momentum} in the
form of
\beq
   {\bf J} =    \bfxi \times \bfeta  + \frac{\zeta^2}{8m^2}\Big[\bfeta^2 \bfxi \times \bfeta
                +   (\bfeta \cdot \bfkappa)(\bfeta \times
                \bfkappa)\Big]
           =   \Big[  c-\frac{1}{4} b^2 \Big] L^4 \bfepsilon_2 + \CO\Big(\frac{1}{\sqrt{\rho}}\Big)\,.
\eeq
 As can be shown from
(\ref{Correction1}) and (\ref{Correction2}), correction terms to the
mass in new type black holes are given by
\[ \Delta_{R} = -\frac{\zeta^2}{2}\frac{L^4 b}{4}\sqrt{\rho} =
-\frac{b}{2L^2}\sqrt{\rho}\,, \qquad
\Delta_{K} = \frac{\zeta^4}{m^2}\frac{L^8 b}{64}\sqrt{\rho}=
\frac{b}{2L^2}\sqrt{\rho}\,. \]
Note that the total contribution of correction terms to mass
vanishes. Finally, one can see that mass of new type black holes is
given by
\beq M= \frac{1}{4\zeta GL}  \bigg[ -\frac{\zeta^2}{2}\Delta J^2
\bigg] = \frac{b^2-4c}{16G}\,,\eeq
which is consistent with  the result by  $AdS/CFT$ correspondence
and satisfies the simple form of the first law of black hole
thermodynamics as~\cite{Nam:2010dd}  (See~\cite{Giribet:2010ed} for the extension  to the the case of rotating new type black holes)
\[
  dM = TdS_{BH}\,.
\]
This shows us that $AdS/CFT$ correspondence may be used as a guiding
principle to obtain physical quantities for gravity. As we have
already seen in the above results, there is no non-linear effect in
the mass correction terms. The non-linear contribution comes only
from the {\it super angular momentum}.

\section{Conclusion}
In this paper we have obtained mass and angular momentum, as conserved charges, of several black holes  in the so-called new massive gravity(NMG) in three dimension and confirmed the validity of first law of black hole thermodynamics and/or $AdS/CFT$ correspondence in this case.  At first, we obtained the ADT potential for the scalar curvature square term and Ricci tensor square one in arbitrary dimension.  This leads to the ADT potential for the NMG case, which is our main interest. Then we obtained  conserved charges of various black hole solutions, which have at least  two commuting Killing vectors. To obtain conserved charges consistent with the first law of black hole thermodynamics and/or $AdS/CFT$ correspondence, we have used {\it super angular momentum} method in the $SO(1,2)$ reduction approach supplemented by the ADT formalism, which is developed by Clement~\cite{Clement:1994}\cite{Clement:1994gr}\cite{Clement:2007}. In the following, several comments are in order.

It has been  well known that  conserved charges in gravity with higher derivatves are difficult to define. There are   attempts called the ADT formalism to accomplish this goal~\cite{Abbott:82}$\sim$\cite{Deser:2003}, which may be regarded as a natural generalization of the famous ADM formalism. Applying this linear ADT formalism, some charges consistent with the first law of black hole thermodynamics are computed and presented in the Appendix D.  Though this approach is covariant,  this  basically uses the linearized peturbation around the given background and so might be incomplete.

 On the other hand, especially in three dimensional case, there is  the so-called {\it super angular momentum}  method~\cite{Clement:1994}\cite{Clement:1994gr} which is not fully covariant but  may capture   non-linear effects.   The nature of {\it super angular momentum} as  the non-linear  conserved quantity  comes from the fact that it  is obtained by the first intergral of equations of motion without perturbative expansion.  Though   {\it super angular momentum}  leads to correct results in some cases,  it turned out to be inconsistent with the first law of black hole thermodynamics in the case of  warped $AdS_3$ black holes in TMG, and then it was shown that there is a  way to overcome this difficulty by supplementing  {\it super angular momentum} with the ADT  formalism~\cite{Clement:2007}\cite{Clement:2009gq}.  That is to say, by comparing the ADT charges with the linearized   {\it super angular momentum}  a linear  correction term to mass was shown to exist in the {\it super angular momentum} side.   Though there is no {\it a priori} guarantee for the absence of non-linear effects in the correction term, it was sufficient for the purpose.

After the advent of $AdS/CFT$ correspondence there are approaches using boundary theory~\cite{Anninos:2008fx} \cite{Nam:2010dd} \cite{Hohm:2010bs}\cite{Giribet:2010ed} to define the bulk $AdS$ conserved charges. Though  this approach is covariant and can capture non-linear effects, it is applicable only to asymptotically (warped) $AdS$ space.  Furthermore, one may need some additional information to define conserved charges in this approach and it is difficult to apply this method  to some cases, for instance,  to warped $AdS_3$ black holes without {\it a priori} inputs. 

 In this paper we have adopted Clement's  {\it hybrid} formalism  of   {\it super angular momentum} method and ADT approach to obtain mass and angular momentum of various black holes in the NMG case. This {\it hybrid} formalism gives us    results consistent with the first law of black hole thermodynamics in all the cases considered in this paper.  Since $AdS/CFT$ approach leads to the same results with ours in all the cases, our results can be regared as consistent with $AdS/CFT$ correspondence.

In  the present state of things  there are no completely satisfactory {\it covariant} non-linear method to define conserved charges in general.  Several approaches have their own merits and demerits as was commented in the above. It will be very interesting to improve this situation  and resolve some issues, for example, whether or not the non-linear effects exist generically for the mass correction term in the {\it hybrid} formalism.

\section*{\center Acknowledgements}
This work was supported by the National Research Foundation of Korea(NRF) grant funded by the Korea government(MEST) through the Center for Quantum Spacetime(CQUeST) of Sogang University with grant number 2005-0049409.
S.H.Y would like to thank  Prof. Seungjoon Hyun  at Yonsei
University for discussion and encouragement.  S.H.Y was supported in part by
the National Research Foundation of Korea (NRF) grant funded by the
Korea government (MEST) with the grant number 2009-0085995. This
work of S.N and J.D.P was supported by a grant from the Kyung Hee
University in 2009(KHU-20100130). S.N was also supported by the
National Research Foundation of Korea(NRF) grant funded by the
Korean government(MEST)(No. 2009-0063068).

\newpage
\appendix
\begin{center}
 \large{\bf APPENDICES}
\end{center}
In this appendix, some calculational details are given.  In section
$A$, we present some details for  ADT charge computation and in
section $B$ give various formulae for vector calculation. In section
$C$, some details about vector representation of the ADT potential
are given and the ADT potential in Schwarzschild coordinates are
given in the appendix D.
  \renewcommand{\theequation}{A.\arabic{equation}}
  \setcounter{equation}{0}
%
\section{The Formulae for the ADT Potential Computation}
Some Killing vector properties:\\
Our curvature convention is
\[ [\nabla_{\alpha}, \nabla_{\beta} ] V^{\mu}= - R^{\mu}_{~\nu\alpha\beta}V^{\nu}\,,\]
which implies  for a scalar $S$
\[
 [\nabla_{\mu}, \nabla^2]S = [\nabla_{\mu}, \nabla_{\nu}]\nabla^{\nu}S = -R_{\mu\nu}\nabla^{\nu}S\,.
\]
Covariant derivatives of Killing vectors, $\xi$, satisfy
\beq
 \nabla_{\mu}\nabla_{\nu}\xi_{\alpha} = -R^{\beta}_{~\mu\alpha\nu}\xi_{\beta}\,, \qquad \nabla^2\xi^{\mu} = -R^{\mu}_{~\nu}\xi^{\nu}\,.
\eeq
and
\bea \CL_{\xi}R &=&  \xi^{\mu}\nabla_{\mu}R=0\,, \nn \\
 \CL_{\xi} \nabla_{\mu}R &=&  \xi^{\alpha}\nabla_{\alpha}\nabla_{\mu}R  + \nabla_{\alpha}R\nabla_{\mu}\xi^{\alpha} =0\,, \nn \\
 \CL_{\xi}R_{\mu\nu} &=& \xi^{\alpha}\nabla_{\alpha}R_{\mu\nu} + R_{\mu\alpha}\nabla_{\nu}\xi^{\alpha} + R_{\nu\alpha}\nabla_{\mu}\xi^{\alpha} =0\,. \nn
 \eea
Bianchi identity and the Killing property lead to the following identity
\beq
 R_{\mu\alpha\nu\beta}\nabla^{\alpha}\xi^{\beta} = \half R_{\mu\nu\alpha\beta}\nabla^{\alpha}\xi^{\beta}\,.
\eeq
The variation of some quantities
\bea
 \delta \Gamma^{\alpha}_{\mu\nu} &=& \half \Big(\nabla_{\mu}h^{\alpha}_{\nu} + \nabla_{\nu}h^{\alpha}_{\mu} -\nabla^{\alpha}h_{\mu\nu}\Big)\,, \nn \\
 \nabla^{\mu}\delta \Gamma^{\alpha}_{\mu\nu} &= & -g^{\alpha\beta}\delta R_{\nu\beta} -\half \nabla_{\nu}\nabla^{\alpha}h+\nabla^{\beta}\nabla_{\nu}h^{\alpha}_{\beta}\,, \nn
 \\
 \delta (\nabla^2 R) &=& \nabla^2\delta R - h_{\alpha\beta}\nabla^{\alpha}\nabla^{\beta}R -g^{\alpha\beta}\delta \Gamma_{\alpha\beta}^{\nu}\nabla_{\nu}R \nn \\
&=&  \nabla^2\delta R - \nabla^{\alpha}(h_{\alpha\beta}\nabla^{\beta}R) +\half \nabla^{\alpha}h\nabla_{\alpha}R\,, \nn \\
\delta (\nabla_{\mu}\nabla_{\nu}R) &=& \nabla_{\mu}\nabla_{\nu}\delta R - \delta \Gamma^{\alpha}_{\mu\nu}\nabla_{\alpha}R\,, \nn \\
\delta (\nabla^2R_{\mu\nu}) &=& \nabla^2\delta R_{\mu\nu} - h_{\alpha\beta}\nabla^{\alpha}\nabla^{\beta}R_{\mu\nu}  -g^{\alpha\beta}\delta \Gamma_{\alpha\beta}^{\nu}\nabla_{\nu}R_{\mu\nu} \nn \\
&&-\nabla^{\alpha}(\delta \Gamma^{\beta}_{\alpha\mu}R_{\beta\nu}) - \nabla^{\alpha}(\delta \Gamma^{\beta}_{\alpha\nu}R_{\beta\mu}) -  \delta \Gamma^{\beta}_{\alpha\mu}\nabla^{\alpha}R_{\beta\nu}- \delta \Gamma^{\beta}_{\alpha\nu}\nabla^{\alpha}R_{\beta\mu}\,, \nn \\
\delta (R_{\alpha\beta}R^{\alpha\beta}) 
&=&  2h^{\alpha\beta}R_{\rho\alpha\beta\sigma}R^{\rho\sigma}+ R_{\alpha\beta}\Big[ \nabla^{\alpha}\nabla_{\gamma}h^{\gamma\beta} +  \nabla^{\beta}\nabla_{\gamma}h^{\gamma\alpha}-\nabla^{\alpha}\nabla^{\beta}h-\nabla^2h^{\alpha\beta}\Big]\,. \nn \eea

\noindent Some formulae for $Q^{\mu\nu}_{R_2}$ computation:\\
Each $E^{\mu}_{a}$ $(a=1,2,\cdots,5)$ term is computed separately,
as follows. First,  let us consider $E^{\mu}_{1}$ term. The same
term may be subtracted from and added to $E^{\mu}_{1}$  just for our
calculational convenience. Then,  the $E^{\mu}_{1}$ term becomes
\bea E^{\mu}_{1}
&=&-h^{\mu\alpha}R_{\alpha\rho\sigma\nu}R^{\rho\sigma}\xi^{\nu} + ( R^{\mu}_{~\, \alpha\gamma\nu}+2R^{\mu}_{~\gamma\alpha\nu})\xi^{\nu} R_{\beta}^{\gamma}h^{\alpha\beta}   + \xi^{\mu}R_{\gamma\alpha}R^{\gamma}_{\beta}h^{\alpha\beta} \nn \\
&& +  A^{\mu}_{1} -\half \xi^{\mu}R^{\alpha\beta}\nabla^{2}h_{\alpha\beta}  +A^{\mu}_{2}  - R^{\mu\alpha}\xi^{\nu}\Big(\nabla^{\gamma}\nabla_{\alpha}h_{\gamma\nu}    -\nabla^2h_{\alpha\nu}\Big)      - 2 \xi^{\nu}\delta \Gamma^{\beta}_{\alpha\nu}\nabla^{\alpha}R_{\beta}^{\mu}\nn \\
&& + A^{\mu}_{3}  + 2\xi^{\gamma}R_{\gamma\alpha\beta\nu}\nabla^{\nu}\nabla^{[\mu}h^{\alpha]}  +  \half \xi^{\mu}R^{\alpha\beta}\nabla_{\alpha}\nabla_{\beta}h\,, \nn
\eea
where
\bea  A^{\mu}_{1} &\equiv &    - R^{\alpha\beta} \xi^{\nu} \Big( \nabla_{\nu}\nabla^{\mu}h_{\alpha\beta}  -\nabla_{\nu}\nabla_{\alpha}h^{\mu}_{\beta}\Big)  -\xi^{\mu}R^{\alpha\beta}\Big(\nabla^{\gamma}\nabla_{\alpha}h_{\gamma\beta}-\nabla^2h_{\alpha\beta}\Big) \nn \\
&=& \nabla_{\nu}\bigg[ 2R^{\alpha\beta}\xi^{[\mu}\nabla^{\nu]}h_{\alpha\beta}-2R^{\alpha\beta}\xi^{\mu}\nabla_{\alpha}h^{\nu]}_{\beta}\bigg]  \nn \\
&& + \xi^{\nu}\nabla_{\nu}R^{\alpha\beta}\Big(\nabla^{\mu}h_{\alpha\beta}-\nabla_{\alpha}h^{\mu}_{\beta}\Big) + \nabla_{\nu}(\xi^{\mu}R^{\alpha\beta})\Big(\nabla_{\alpha}h^{\nu}_{\beta}-\nabla^{\nu}h_{\alpha\beta}\Big)\,, \nn  \\
&& \nn \\
 A^{\mu}_{2} &\equiv &    - R^{\alpha\beta} \xi^{\nu} \Big( \nabla_{\beta}\nabla_{\alpha}h^{\mu}_{\nu} -\nabla_{\beta}\nabla^{\mu}h_{\alpha\nu} \Big)  +R^{\mu\alpha}\xi^{\nu}\Big(\nabla^{\gamma}\nabla_{\alpha}h_{\gamma\nu}-\nabla^2h_{\alpha\nu}\Big) \nn \\
 &=& \nabla_{\nu}\bigg[ 2\xi^{\beta}R^{\alpha[\mu}\nabla_{\alpha}h^{\nu]}_{\beta} - 2\xi^{\beta} R^{\alpha[\mu} \nabla^{\nu]}h_{\alpha\beta} \bigg]  \nn \\
 &&+ \nabla_{\nu}(\xi^{\beta}R^{\mu\alpha})\Big(\nabla^{\nu}h_{\alpha\beta}-\nabla_{\alpha}h^{\nu}_{\beta}\Big) + \nabla_{\nu}(\xi^{\beta}R^{\nu\alpha})\Big(\nabla^{\mu}h_{\alpha\beta}-\nabla_{\alpha}h^{\mu}_{\beta}\Big)\,, \nn \\ && \nn \\
 A^{\mu}_{3} &\equiv &
-2R^{\mu\alpha\beta}_{~~~\,\,\nu}\xi^{\nu}\delta R_{\alpha\beta} - 2\xi^{\gamma}R_{\gamma\alpha\beta\nu}\nabla^{\nu}\nabla^{[\mu}h^{\alpha]}\nn \\
&=& \nabla_{\nu} \bigg[ R^{\mu\nu}_{~~\, \alpha\beta} \xi^{\alpha} \nabla_{\gamma}h^{\gamma\beta}-2\xi^{\gamma}R_{\gamma}^{~\alpha\beta[\mu}\nabla_{\alpha}h^{\nu]}_{\beta} + 2\xi^{\gamma}R_{\gamma}^{~\alpha\beta[\mu}\nabla^{\nu]}h_{\alpha\beta}\bigg]  \nn  \\     \nn
&&+ \nabla_{\nu}(\xi^{\gamma}R_{\gamma}^{~\alpha\beta\mu}) \Big(\nabla_{\alpha}h^{\nu}_{\beta} - \nabla^{\nu}h_{\alpha\beta}\Big)  +   \nabla_{\nu}(\xi^{\gamma}R_{\gamma}^{~\alpha\beta\nu}) \Big(\nabla^{\mu}h_{\alpha\beta}  - \nabla_{\alpha}h^{\mu}_{\beta} \Big) \nn \\
&& + \xi^{\nu}(\nabla_{\nu}R^{\mu}_{\beta} -\nabla_{\beta} R^{\mu}_{\nu})\nabla_{\alpha}h^{\alpha\beta} -R^{\mu}_{~\rho\sigma\beta}\nabla^{\rho}\xi^{\sigma} \nabla_{\alpha}h^{\alpha\beta} \nn  \\
&& -R^{\mu}_{~\gamma\alpha\nu}\xi^{\nu}R^{\gamma}_{\beta}h^{\alpha\beta} -R^{\mu}_{~\rho\sigma\nu}\xi^{\nu} R^{\sigma\alpha\beta\rho}h_{\alpha\beta}  -R^{\mu}_{~\alpha\beta\nu}\xi^{\nu}\nabla^{\alpha}\nabla^{\beta}h\,. \nn
\eea
For $A^{\mu}_{1}$ and $A^{\mu}_{2}$, one can see that
\bea
&& A^{\mu}_{1} + A^{\mu}_{2} - 2 \xi^{\nu}\delta \Gamma^{\beta}_{\alpha\nu}\nabla^{\alpha}R_{\beta}^{\mu}  \nn \\
&& = \nabla_{\nu}\bigg[ 2\xi^{\alpha}R^{\beta[\mu}\nabla_{\beta}h^{\nu]}_{\alpha}-2\xi^{\alpha}R^{\beta[\mu}\nabla^{\nu]}h_{\alpha\beta}+2R^{\alpha\beta}\xi^{[\mu}\nabla^{\nu]}h_{\alpha\beta}-2R^{\alpha\beta}\xi^{[\mu}\nabla_{\alpha}h^{\nu]}_{\beta} +2h^{\alpha\beta}\nabla_{\alpha}(\xi^{[\mu}R^{\nu]}_{\beta})\bigg]  \nn \\
&&~~~   -\half \xi^{\nu}\nabla^{\alpha}R(\nabla^{\mu}h_{\alpha\nu}-\nabla_{\alpha}h^{\mu}_{\nu}) - R_{\alpha\nu}\nabla^{\nu}\xi^{\beta}(\nabla^{\mu}h^{\alpha}_{\beta}-\nabla^{\alpha}h^{\mu}_{\beta})   \nn \\
&&~~~+ \xi^{\gamma}\nabla_{\gamma}R^{\alpha\beta}(\nabla^{\mu}h_{\alpha\beta}-\nabla_{\alpha}h^{\mu}_{\beta})  -\nabla_{\nu}(\xi^{\mu}R^{\alpha\beta})\nabla^{\nu}h_{\alpha\beta}      -h^{\alpha\beta}\nabla_{\nu}\nabla_{\beta}(\xi^{\mu}R^{\nu}_{\alpha}-\xi^{\nu}R^{\mu}_{\alpha}) \,.  \nn  \eea
For $A^{\mu}_{3}$, one may note that the following identities
\bea \nabla_{\nu}(\xi^{\gamma}R_{\gamma}^{~\alpha\beta\mu}) \Big(\nabla_{\alpha}h^{\nu}_{\beta} - \nabla^{\nu}h_{\alpha\beta}\Big)  &=& \nabla_{\nu}\Big( h_{\alpha\beta}R_{\gamma}^{~\alpha\beta\nu}\nabla^{\mu}\xi^{\gamma}\Big) - h^{\alpha\beta}\nabla_{\alpha}R_{\beta\gamma}\nabla^{\mu}\xi^{\gamma}  - h^{\alpha\beta}\nabla_{\gamma}R_{\alpha\beta}\nabla^{\gamma}\xi^{\mu}   \nn \\
&&+ R^{\mu}_{~\rho\sigma\nu}\xi^{\nu}R^{\sigma\alpha\beta\rho}h_{\alpha\beta} + R^{\mu}_{~\alpha\nu\beta}\nabla^{\nu}\xi^{\gamma}(\nabla^{\alpha}h^{\beta}_{\gamma} - \nabla^{\beta}h^{\alpha}_{\gamma})\,,  \nn  \\ && \nn \\
\nabla_{\nu}(\xi^{\gamma}R_{\gamma}^{~\alpha\beta\nu}) \Big(\nabla^{\mu}h_{\alpha\beta}  - \nabla_{\alpha}h^{\mu}_{\beta} \Big)  &=&  \xi^{\gamma}(\nabla_{\alpha}R_{\beta\gamma}-\nabla_{\gamma}R_{\alpha\beta})(\nabla^{\mu}h^{\alpha\beta} - \nabla^{\alpha}h^{\mu\beta}) -  R_{\gamma\alpha\beta\nu} \nabla^{\nu} \xi^{\gamma} \nabla^{\alpha}h^{\mu\beta}\,, \nn \\ && \nn \\
h^{\alpha\beta}\nabla_{\nu}\nabla_{\beta}(\xi^{\mu}R^{\nu}_{\alpha}-\xi^{\nu}R^{\mu}_{\alpha})
&=&  h^{\alpha\beta}\Big(\nabla_{\alpha}R^{\mu}_{\nu}\nabla_{\beta}\xi^{\nu} + \half \nabla_{\alpha}R\nabla_{\beta}\xi^{\mu}\Big)    -h^{\alpha\beta}\xi^{\nu}\Big(
R^{\mu}_{\alpha}R_{\beta\nu} -R^{\mu}_{~\alpha\gamma\nu}R^{\gamma}_{\beta}\Big)  \nn \\
&&+ \xi^{\mu} h^{\alpha\beta}\Big(R_{\gamma\alpha}R^{\gamma}_{\beta} + R_{\alpha\rho\sigma\beta}R^{\rho\sigma} +\half \nabla_{\alpha}\nabla_{\beta}R\Big)\,. \nn \eea
Collecting various expression in the above, one can obtains $E^{\mu}_{1}$.

For the calculation of $E^{\mu}_{2}$ term, one may note that
\[ \nabla^2\nabla_{\nu}Q^{\mu\nu}_R = \nabla_{\nu}\Big[\nabla^2Q^{\mu\nu}+R^{\mu\nu}_{~~\alpha\beta}Q^{\alpha\beta}+2Q^{\alpha[\mu}R^{\nu]}_{\alpha}\Big] + R^{\mu}_{\alpha}\nabla_{\nu}Q^{\alpha\nu}\,, \]
which leads to
\bea
\xi^{\nu}\nabla^2\CI^{\mu}_{~\nu} & =&\nabla^2(\CI^{\mu}_{~\nu}\xi^{\nu}) -\CI^{\mu}_{~\nu}\nabla^2\xi^{\nu} -2\nabla_{\nu}\CI^{\mu}_{~\alpha}\nabla^{\nu}\xi^{\alpha} \nn \\
&=&\nabla^2(\CI^{\mu}_{~\nu}\xi^{\nu}) -\nabla_{\nu}(\delta R \nabla^{[\mu}\xi^{\nu]})   -2g^{\mu\rho}\nabla_{\alpha}\delta R_{\rho\nu} \nabla^{\alpha}\xi^{\nu}
+\CI^{\mu}_{~\alpha}R^{\alpha}_{\nu}\xi^{\nu}  + \delta RR^{\mu}_{\nu}\xi^{\nu}  \nn \\
&=& \nabla_{\nu}\bigg[\nabla^2Q^{\mu\nu}_R + R^{\mu\nu}_{~~\alpha\beta}Q^{\alpha\beta}_R + 2 R_{\alpha}^{[\mu} Q^{\nu]\alpha}_R -\delta R\nabla^{[\mu}\xi^{\nu]}\bigg]    \nn \\
&&  -2g^{\mu\rho}\nabla_{\nu}\delta R_{\rho\alpha}\nabla^{\nu}\xi^{\alpha}  +\nabla^2\Big(h^{\mu\beta}R_{\beta\nu}\xi^{\nu}+\half \xi^{\mu}R^{\rho\sigma}h_{\rho\sigma}-\half R^{\mu}_{\nu}\xi^{\nu}h\Big)  \nn \\
&&+(\CI^{\mu}_{~\alpha}R^{\alpha}_{\nu} + R^{\mu}_{\alpha}\CI^{\alpha}_{~\nu})\xi^{\nu} + \delta RR^{\mu}_{\nu}\xi^{\nu} - R^{\mu}_{\alpha}\Big(h^{\alpha\beta}R_{\beta\nu}\xi^{\nu}+\half \xi^{\alpha}R^{\rho\sigma}h_{\rho\sigma}-\half R^{\alpha}_{\nu}\xi^{\nu}h\Big) \nn
\,.\eea
Noting that   the following  formula
\bea
   -2g^{\mu\rho}\nabla_{\nu}\delta R_{\rho\alpha}\nabla^{\nu}\xi^{\alpha}
   &=&  -2\nabla_{\nu}\bigg[\nabla^{\alpha}\xi^{\beta} \nabla_{\alpha}\nabla^{[\mu}h^{\nu]}_{\beta}\bigg] + \half R^{\mu}_{~\nu\alpha\beta}\nabla^{\alpha}\xi^{\beta} \nabla^{\nu}h - 2\xi^{\gamma}R_{\gamma\alpha\beta\nu}\nabla^{\nu}\nabla^{[\mu}h^{\alpha]\beta}\nn \\
   && +R_{\alpha\nu}\nabla^{\nu}\xi^{\beta}
   (\nabla^{\mu}h^{\alpha}_{\beta}-\nabla^{\alpha}h^{\mu}_{\beta})   - \nabla_{\gamma}(h^{\mu\alpha}R_{\alpha\nu} +R^{\mu}_{~\alpha\beta\nu}h^{\alpha\beta}) \nabla^{\gamma}\xi^{\nu} \nn \\
   &&+ R^{\mu}_{~\alpha\nu\beta}\nabla^{\nu}\xi^{\gamma}(\nabla^{\beta}h^{\alpha}_{\gamma} - \nabla^{\alpha}h^{\beta}_{\gamma})\nn   \\
   &&  + R^{\gamma}_{~\alpha\beta\nu} \nabla^{\beta}\xi^{\alpha}(\nabla^{\mu}h^{\nu}_{\gamma}-\nabla_{\nu}h^{\mu}_{\gamma})   -\half R^{\mu}_{~\nu\rho\sigma} \nabla_{\alpha}h^{\alpha\beta}\nabla^{\rho}\xi^{\sigma}\,. \nn
\eea
and  the formula containing the $F_{\mu\nu}$ term in~(\ref{Fterm})
 \bea  &&  F^{\mu}_{~\nu}\xi^{\nu}  +R_{\alpha\nu}\xi^{\nu}\nabla^2h^{\mu\alpha} + \CI^{\mu}_{~\alpha}R^{\alpha}_{\nu}\xi^{\nu}  +R^{\mu}_{\alpha}\CI^{\alpha}_{~\nu}\xi^{\nu} \nn \\
 &=&  \nabla_{\nu}\bigg[ -2\xi^{\alpha}R_{\alpha}^{[\mu}\nabla_{\beta}h^{\nu]\beta}+\xi^{\alpha}R_{\alpha}^{[\mu}\nabla^{\nu]}h\bigg] +  \nabla_{\beta}(\xi^{\alpha}R_{\alpha}^{\mu})\nabla_{\alpha}h^{\alpha\beta}    \nn \\
&&+ R^{\mu}_{~\rho\sigma\alpha}h^{\rho\sigma}R^{\alpha}_{\nu}\xi^{\nu} +  h^{\mu\alpha}R_{\alpha\beta}R^{\beta}_{\nu}\xi^{\nu} +   R^{\mu}_{\nu}\xi^{\nu}R^{\alpha\beta}h_{\alpha\beta} \nn \\
&& + R^{\mu\alpha}\xi^{\nu}(\nabla^{\gamma}\nabla_{\alpha}h_{\gamma\nu}-\nabla^2h_{\alpha\nu}) - \half R^{\mu\alpha}\xi^{\nu}\nabla_{\nu}\nabla_{\alpha} h + \half R^{\mu}_{\nu}\xi^{\nu}\nabla^2h -\half \nabla_{\nu}(\xi^{\alpha}R_{\alpha}^{\mu}) \nabla^{\nu}h \,, \nn      \eea
one can obtain $E_2$ expression.

%

Formulae for remaining $E^{\mu}_a$ are
\bea
E^{\mu}_{3}
&=& \nabla_{\nu}\bigg[ 2\xi^{\beta}h^{\alpha[\mu}\nabla_{\alpha}R^{\nu]}_{\beta}\bigg] -h^{\mu\alpha}\xi^{\nu} \Big(R_{\alpha\rho\sigma\nu}R^{\rho\sigma} +\half \nabla_{\alpha}\nabla_{\nu}R + R_{\alpha\beta}R^{\beta}_{\nu}\Big) \nn \\
&& + h^{\alpha\beta}\nabla_{\alpha}R^{\mu}_{\nu}\nabla_{\beta}\xi^{\nu} - \xi^{\nu}\nabla^{\alpha}R^{\beta}_{\nu}(\nabla_{\alpha}h^{\mu}_{\beta}+\nabla^{\mu}h_{\alpha\beta} )\,. \nn \\ && \nn \\
E^{\mu}_{4}
&=& \nabla_{\nu}\bigg[ -\xi^{[\mu}h^{\nu]\alpha}\nabla_{\alpha}R \bigg]  + \half h^{\alpha\beta}\nabla_{\alpha}R \nabla_{\beta}\xi^{\mu} - \half h^{\mu\alpha}\xi^{\nu}\nabla_{\alpha}\nabla_{\nu}R + \half \xi^{\nu}\nabla^{\alpha}R(h^{\mu}_{\alpha\nu} - \nabla_{\alpha}h^{\mu}_{\nu})\,.\nn   \\ && \nn \\
E^{\mu}_{5} &\equiv& \half \xi^{\nu} \nabla_{\alpha}h\nabla^{\alpha}R^{\mu}_{\nu}  + \frac{1}{4}\xi^{\mu} \nabla_{\alpha}h\nabla^{\alpha}R\,.\nn
\eea
%

 \section{Identities and Useful Formulae for Vector Representation}
  \renewcommand{\theequation}{B.\arabic{equation}}
  \setcounter{equation}{0}
Ricci tensor and scalar curvature are given by
\bea
R^{a}_{b} & =& -\frac{\zeta^2}{2}\Big[(UU')' + \langle{\bf L}\rangle'\Big]^{a}_{b} \,, \qquad R^{\rho}_{\rho} = -\frac{\zeta^2}{2}\Big[2(UU')' - {\bf X}'^2 \Big]\,, \nn \\
R &=& -\frac{\zeta^2}{2}\Big[ 4(UU')' - {\bf X}'^2\Big] = -\frac{\zeta^2}{2}\Big[4{\bf X}\cdot {\bf X}'' + 3{\bf X}'^2\Big]\,. \eea
Covariant derivative of Ricci tensors
\bea
\nabla_{\rho}R^{\rho}_{\rho} &=& (R^{\rho}_{\rho})' = -\zeta^2\Big[(UU')''-{\bf X}'\cdot {\bf X}'' \Big]\,, \nn
\\
\nabla_{\rho}R^{b}_{a} &=& (R^{b}_{a})' + \half (\lambda^{-1}\lambda')^{b}_{c}R^{c}_{a} - \half R^{b}_{c}(\lambda^{-1}\lambda')^{c}_{a} \nn \\
&=&-\frac{\zeta^2}{2} \Big[(UU')'' + \langle {\bf L}'' \rangle + \half(\lambda^{-1}\lambda' \langle {\bf L}' \rangle - \langle {\bf L}' \rangle\lambda^{-1}\lambda' ) \Big]\,,
\nn
\\
\nabla_{a}R^{b}_{\rho} &=& R^{\rho}_{\rho}\Gamma^{b}_{\rho a} -
R^{b}_{c}\Gamma^{c}_{\rho a} = \half
R^{\rho}_{\rho}(\lambda^{-1}\lambda')^{b}_{a} -\half
R^{b}_{c}(\lambda^{-1}\lambda')^{c}_{a}= \frac{\zeta^2}{4}\Big[
(-{\bf X}\cdot {\bf X}'')  +\langle {\bf L}' \rangle
\Big]\lambda^{-1}\lambda'\,, \nn
\\
\nabla_{a}R^{\rho}_{b} &=& \frac{\zeta^4}{4} U^2 \lambda'
\Big[-({\bf X}\cdot {\bf X}'') + \langle {\bf L}' \rangle \Big] \,.
\eea
Covariant derivative of metric deviation
\bea
\nabla_{b}h^{a}_{\rho} &=& -\frac{\delta U}{U}(\lambda^{-1}\lambda')^{a}_{b} -\half(\lambda^{-1}\delta \lambda\lambda^{-1}\lambda')^{a}_{b}\,, \qquad \nabla_{c}h^{a}_{b} =0\,, \qquad \nabla_{a}h^{\rho}_{\rho}=0\,, \nn \\
\nabla_{a}h^{\rho}_{b} &=& -\frac{\zeta^2}{2}U^2\Big( \lambda'\lambda^{-1}\delta \lambda  +2\lambda'\frac{\delta U}{U}\Big)_{ab}   \,, \nn \\
\nabla_{\rho} h^{a}_{b} &=& (h^{a}_{b})' + \half
(\lambda^{-1}\lambda'\lambda^{-1}\delta \lambda)^{a}_{b} - \half
(\lambda^{-1}\delta \lambda\lambda^{-1}\lambda')^{a}_{b}\,,   \qquad
\nabla_{\rho}h^{\rho}_{\rho} = (h^{\rho}_{\rho})'\,,   \qquad
\nabla_{\rho}h^{\rho}_{a}=\nabla_{\rho}h^{a}_{\rho}=0\,,  \nn \eea
which imply
\bea
\nabla_{\mu}h^{\mu}_{\rho} &=&(h^{\rho}_{\rho})' -2\frac{U'\delta U}{U^2} -\half \Tr(\lambda^{-1}\lambda'\lambda^{-1}\delta \lambda)=-\frac{1}{U^2}\Big[{\bf X}'\cdot \delta {\bf X} + 2{\bf X} \cdot \delta {\bf X}'\Big]\,,  \nn
\\
\nabla_{\mu}h^{\mu}_{a} &=&0\,, \qquad
 2\nabla_{[\rho}h^{a}_{b]}  =  (\lambda^{-1}\delta \lambda)'^{a}_{~b} + \half (\lambda^{-1}\lambda'\lambda^{-1}\delta \lambda)^{a}_{b} +\frac{\delta U}{U}(\lambda^{-1}\lambda')^{a}_{b} \,.  \nn \eea
Useful vector identities:
\bea
 {\bf L}^2 &=& U^2(U'^2-{\bf X}'^2)\,,
 \\
 {\bf L}\cdot {\bf \Sigma} &=& U^2(U'\delta U - {\bf X}'\cdot \delta {\bf X})=U^2(-U\delta U' + {\bf X}\cdot \delta {\bf X}')\,, \nn
 \\
 {\bf L}'\cdot {\bf \Sigma} &=& U\delta U {\bf X}\cdot {\bf X}''- U^2  {\bf X}''\cdot \delta {\bf X}\,, \nn
 \\
 {\bf L}\cdot {\bf \Sigma}' &=&  U\delta U\, {\bf X}'^2   - U^2  {\bf X}'\cdot \delta {\bf X}'-U U'({\bf X}'\cdot \delta {\bf X} - {\bf X}\cdot \delta {\bf X}')\,, \nn
 \\
 {\bf L}'\cdot {\bf \Sigma}' &=& U\delta U {\bf X}'\cdot {\bf X}'' -UU'{\bf X}''\cdot \delta {\bf X} + ({\bf X}\cdot {\bf X}''){\bf X}\cdot \delta {\bf X}' -U^2{\bf X}''\cdot \delta {\bf X}' \nn
 \\ {\bf L}\cdot \delta {\bf L} &=& UU'(U\delta U)' - U\delta U\, {\bf X}'^2 -U^2{\bf X}'\cdot \delta {\bf X}'\,, \nn
 \\
 {\bf L}\cdot {\bf L}' &=& UU'{\bf X}\cdot {\bf X}'' -U^2{\bf X}'\cdot{\bf X}''\,, \nn
 \\
  {\bf L}'\cdot \delta {\bf L} &=& UU'{\bf X}''\cdot \delta {\bf X} + ({\bf X}\cdot {\bf X}'')\, {\bf X}\cdot \delta {\bf X}'- U\delta U {\bf X}'\cdot {\bf X}'' -U^2{\bf X}''\cdot \delta {\bf X}'\,, \nn
 \\
  {\bf L}\times {\bf \Sigma} &=& \half  U^2(\delta {\bf L} - {\bf \Sigma}') +UU'{\bf \Sigma}-U\delta U{\bf L}\,, \nn
 \\
   {\bf L}'\times {\bf \Sigma} &=& U^2\delta {\bf X} \times {\bf X}'' +({\bf X}\cdot {\bf X}''){\bf \Sigma}-U\delta U{\bf L}'\,, \nn
 \\
  {\bf L}\times {\bf \Sigma}' &=&UU'\delta {\bf L} -(U\delta U)'{\bf L} -  U^2 {\bf X}' \times \delta  {\bf X}' +{\bf X}'^2{\bf \Sigma}\,, \nn
 \\
  {\bf L}'\times   {\bf \Sigma}' &=& -({\bf X}\cdot \delta {\bf X}')\, {\bf L}' + U^2\delta {\bf X}'\times {\bf X}''  + UU'\delta {\bf X}\times {\bf X}''  - U\delta U{\bf X}'\times {\bf X}'' - ({\bf X}''\cdot \delta {\bf X})\, {\bf L} \nn
  \\
  && + ({\bf X}\cdot {\bf X}'') \, {\bf X}\times \delta {\bf X}' + ({\bf X}'\cdot {\bf X}'')\, {\bf \Sigma}\,, \nn
 \\
  {\bf L}\times {\bf L}' &=&UU'{\bf L}' - ({\bf X}\cdot {\bf X}'')\, {\bf L}   - U^2 {\bf X}'\times {\bf X}''\,, \nn
  \\
 {\bf L}\times \delta {\bf L} &=& (\delta {\bf X}\cdot{\bf X}'  - {\bf X}\cdot \delta {\bf X}'){\bf L} +UU'{\bf \Sigma}' -U^2{\bf X}'\times \delta {\bf X}' -{\bf X}'^2{\bf \Sigma}\,, \nn
  \\
   {\bf L}'\times \delta {\bf L} &=&-({\bf X}\cdot \delta {\bf X}')\, {\bf L}' + U^2\delta {\bf X}'\times {\bf X}''  -UU'\delta {\bf X}\times {\bf X}'' +U\delta U{\bf X}'\times {\bf X}'' + ({\bf X}''\cdot \delta {\bf X})\, {\bf L}\nn
   \\&& +({\bf X}\cdot {\bf X}'') \, {\bf X}\times \delta {\bf X}'-({\bf X}'\cdot {\bf X}'')\, {\bf \Sigma}\,, \nn
\\
 {\bf L}\times \delta {\bf L}' &=& ({\bf X}'\cdot \delta {\bf X})\, {\bf L}' -U^2{\bf X}'\times \delta {\bf X}'' +UU'{\bf X}\times \delta {\bf X}'' -U\delta U {\bf X}'\times {\bf X}''   - ({\bf X}\cdot \delta {\bf X}'')\, {\bf L}   \nn
 \\
 && -({\bf X}\cdot {\bf X}'')\, \delta {\bf X} \times {\bf X}'    -  ({\bf X}'\cdot {\bf X}'')\, {\bf \Sigma}\,. \nn
\eea
Note that some vector quantities may be written in terms of other ones as follows
\bea
{\bf L}\times {\bf \Sigma}' &=& {\bf L}\times \delta {\bf L} +UU'(\delta {\bf L}- {\bf \Sigma}) +2({\bf X}'^2\, {\bf \Sigma} -{\bf X}'\cdot \delta {\bf X}\, {\bf L})\,, \label{VIDa}
\\
{\bf L}\cdot {\bf \Sigma}' &=& {\bf L}\cdot \delta {\bf L}  +2(U\delta U\, {\bf X}'^2 - UU'{\bf X}'\cdot \delta {\bf X})\,, \nn \\
{\bf L}'\cdot  {\bf \Sigma}' &=& {\bf L}'\cdot \delta {\bf L} +2\bigg[\frac{U'}{U}{\bf L}'\cdot {\bf \Sigma} -\frac{\delta U}{U}{\bf L}\cdot {\bf L}'\bigg]\,,  \nn \\
{\bf L}'\times {\bf \Sigma}' &=& {\bf L}'\times \delta {\bf L}+2\bigg[\frac{\delta  U}{U}{\bf L}\times {\bf L}' + \frac{U'}{U} {\bf L}'\times {\bf \Sigma} +  \frac{{\bf L}'\cdot {\bf \Sigma}}{U^2}  {\bf L} -\frac{{\bf L}\cdot {\bf L}'}{U^2}{\bf \Sigma}\bigg]\,,  \nn \\
{\bf L}'\cdot ({\bf L} \times {\bf \Sigma}) &=& 0\,.  \nn
\eea
Conjugation by two dimensional metric $\lambda$:\\
Note that for any vector ${\bf A}$  one obtains
\beq \lambda \langle {\bf A} \rangle \lambda^{-1}= \tau^0 \bigg[ \langle {\bf A} \rangle  -\frac{2}{U^2}{\bf X}\cdot {\bf A}\langle {\bf X} \rangle\bigg]\tau^0  \,, \eeq
which leads to
\bea
\lambda \langle {\bf X} \rangle \lambda^{-1} &=& -\tau^0 \langle {\bf X} \rangle \tau^0\,,  \qquad \lambda \langle {\bf X}' \rangle \lambda^{-1} = \tau^0\bigg[-\frac{2U'}{U}  \langle {\bf X} \rangle   + \langle {\bf X}' \rangle \bigg]\tau^0\,, \nn
\\
 \lambda \langle {\bf L} \rangle  \lambda^{-1} &=& \tau^0 \langle {\bf L} \rangle   \tau^0\,, \qquad
 \lambda \langle {\bf L}' \rangle  \lambda^{-1} = \tau^0 \langle {\bf L}' \rangle  \tau^0\,, \nn
 \\
 \lambda \langle {\bf L}'' \rangle \lambda^{-1} &=& \tau^0\bigg[ \langle {\bf X}\times {\bf X}''' - {\bf X}'\times {\bf X}'' \rangle + \frac{2U'}{U}\langle {\bf L}' \rangle - \frac{2{\bf X}\cdot{\bf X}''}{U^2} \langle {\bf L} \rangle \bigg] \tau^0 \nn
 \\ &=& \tau^0\bigg\langle {\bf L}'' + \frac{2}{U^2} {\bf L}\times{\bf L}' \bigg\rangle \tau^0\,, \nn
 \\
  \lambda \langle {\bf \Sigma} \rangle  \lambda^{-1} &=& \tau^0 \langle {\bf \Sigma} \rangle   \tau^0\,, \nn
  \\
 \lambda \langle {\bf \Sigma}' \rangle  \lambda^{-1} &=& \tau^0\bigg[-\frac{2}{U^2}{\bf L}\cdot \delta {\bf X} \langle {\bf X}\rangle+\langle {\bf \Sigma}' \rangle \bigg] \tau^0= \tau^0\bigg[\frac{2U' }{U}\langle {\bf \Sigma} \rangle - \frac{2\delta U }{U} \langle {\bf L}\rangle +\langle {\delta \bf L} \rangle \bigg] \tau^0\,, \nn
 \\
 \lambda \langle {\bf \Sigma}'' \rangle  \lambda^{-1} &=& \tau^0\bigg[-\frac{2}{U^2}{\bf X}\cdot  {\bf \Sigma}'' \langle {\bf X}\rangle+\langle {\bf \Sigma}'' \rangle \bigg] \tau^0   \nn
 \\ &=& \tau^0\bigg[  \langle \delta {\bf L}' \rangle  - 2\langle {\bf X}'\times \delta {\bf X}' \rangle    -2\frac{\delta U}{U} \langle {\bf L}' \rangle    -\frac{4}{U^2} {\bf X}\cdot  \delta {\bf X}' \langle {\bf L} \rangle  + 2\frac{{\bf X}\cdot {\bf X}''}{U^2} \langle {\bf \Sigma} \rangle  \nn
 \\&&~~~~~ + 4\frac{U'}{U}\langle {\bf X}\times \delta {\bf X}' \rangle \bigg] \tau^0 \,, \nn
 \\
  \lambda \langle {\bf L}\times {\bf \Sigma} \rangle  \lambda^{-1} &=& -\tau^0\Big[{\bf L}\times {\bf \Sigma}\Big]\tau^0\,, \nn
 \\
 \lambda \langle {\bf L}'\times {\bf \Sigma} \rangle  \lambda^{-1} &=& -\tau^0\Big[{\bf L}'\times {\bf \Sigma}\Big]\tau^0\,, \nn
\eea
and also
\bea
 \half \lambda \Big(\lambda^{-1}\lambda'\langle {\bf L}' \rangle -\langle {\bf L}' \rangle\lambda^{-1}\lambda' \Big)\lambda^{-1} &=& \tau^0\bigg[  \langle {\bf X}' \times {\bf X}'' \rangle  -\frac{U'}{U}\langle {\bf L}' \rangle + \frac{{\bf X}\cdot {\bf X}''}{U^2} \langle {\bf L} \rangle \bigg]\tau^0 \nn
 \\
 &=&\tau^0\bigg\langle    -\frac{1}{U^2} {\bf L}\times {\bf L}' \bigg\rangle \tau^0\,, \nn
 \\
 \half \lambda \Big(\lambda^{-1}\lambda'\langle {\bf L}' \rangle  + \langle {\bf L}' \rangle\lambda^{-1}\lambda' \Big)\lambda^{-1} &=& \frac{U'}{U}{\bf X}\cdot {\bf X}''-{\bf X}'\cdot {\bf X}'' + \tau^0\frac{U'}{U}\langle {\bf L}' \rangle \tau^0 \,, \nn
 \eea
%

\section{Vector Representation of Various Terms in the ADT Potential for NMG}
\renewcommand{\theequation}{C.\arabic{equation}}
  \setcounter{equation}{0}
In this appendix, we present various formulae for the conjugation by two dimensional metric, $\lambda$ or useful formulae to obtain the ADT potential and then present vector representation for various terms in the ADT potential of NMG.

\noindent Useful formulae for the computation of $Q^{\mu\nu}_{R}$:
\bea \lambda (\lambda^{-1}\delta \lambda)'\lambda^{-1} &=&
\frac{1}{U^2}\bigg[(U\delta U' - U'\delta U){\bf 1} +
\tau^0\Big(-2\frac{\delta U}{U} \langle {\bf L}\rangle + \langle
\delta {\bf L}\rangle\Big)\tau^0\bigg]\,,   \\
\half \Tr(\lambda^{-1}\lambda'\lambda^{-1}\delta \lambda)
&=&\frac{2U'}{U^3}\, {\bf X}\cdot \delta {\bf X} -\frac{1}{U^2}\,
{\bf X}'\cdot \delta{\bf X}\,. \eea

\noindent Useful formulae for the computation of $Q^{\mu\nu}_{K}$: We omit the obvious matrix indices in the right hand side of equations.
\bea
 \lambda_{ac} \nabla_{[\rho} h^{c}_{d]} (\lambda^{-1})^{db} &=& \frac{1}{4U^2} ({\bf X}'\cdot \delta {\bf X} + 2 {\bf X}\cdot \delta {\bf X}')  + \frac{1}{8U^2} \tau^0\Big\langle 3\delta {\bf L}  +   {\bf \Sigma}' \Big\rangle\tau^0 \,. \nn \\ && \nn \\
 \lambda'_{ac} \nabla_{[\rho} h^{c}_{d]} (\lambda^{-1})^{db} & =&  \frac{U'}{4U^3}({\bf X}'\cdot \delta {\bf X}+2{\bf X}\cdot \delta {\bf X}') + \frac{1}{8U^4}{\bf L}\cdot (3\delta {\bf L} + {\bf \Sigma}) \nn
\\
&& +  \tau^0\bigg\langle \frac{U'}{8U^3}  (3\delta {\bf L} + {\bf \Sigma}' ) + \frac{1}{4U^4}({\bf X}'\cdot \delta {\bf X} + 2{\bf X}\cdot \delta {\bf X}') {\bf L}
-\frac{1}{8U^4}     {\bf L} \times (3\delta {\bf L} +{\bf \Sigma}' ) \bigg\rangle\tau^0 \,. \nn   \eea
\bea
\lambda'\lambda^{-1}\delta \lambda \lambda^{-1}
  &=&\frac{1}{U^2}\Big( U'\delta U - U\delta U' +  {\bf X}\cdot \delta {\bf X}'\Big)  +\tau^0\bigg\langle 2\frac{\delta U}{U^3}   {\bf L}    -\frac{1}{2U^2}  \delta {\bf L}   + \frac{1}{2U^2}   {\bf \Sigma}'  \bigg\rangle  \tau^0\,.   \nn \\
  && \nn  \\
\lambda \langle{\bf L}'\rangle \lambda^{-1} \lambda' \lambda^{-1}
\delta \lambda \lambda^{-1}  &=& \frac{U'}{U^3}{\bf L}'\cdot {\bf
\Sigma} + \frac{\delta U}{U^3}{\bf L}\cdot {\bf
L}'-\frac{1}{U^4}{\bf L}'\cdot ({\bf L}\times {\bf \Sigma}) \nn
\\ &&+ \tau^0\bigg\langle\Big(\frac{U'\delta U}{U^2}  + \frac{1}{U^4}{\bf L}\cdot {\bf \Sigma} \Big)  {\bf L}'   + \frac{\delta U}{U^3} {\bf L}\times {\bf L}'   -\frac{U'}{U^3}   {\bf L}'\times {\bf \Sigma}  \nn
\\&&~~~~~ + \frac{1}{U^4} {\bf L}'\times ({\bf L} \times {\bf \Sigma})  \bigg\rangle \tau^0\,, \nn \eea
%
%
\bea   \lambda \langle{\bf L}'\rangle \bigg[ (\lambda^{-1}\delta
\lambda)' + \frac{\delta U}{U}\lambda^{-1}\lambda' \bigg]
\lambda^{-1}  &=& \frac{1}{U^2}\Big({\bf L}'\cdot \delta {\bf L} -
\frac{\delta U}{U}{\bf L}\cdot {\bf L}'\Big)\nn
\\
&& +\tau^0\bigg[\frac{\delta U'}{U}\langle {\bf L}' \rangle -\frac{1}{U^2}\Big(\langle {\bf L}'\times \delta {\bf L}\rangle + \frac{\delta U}{U}\langle {\bf L}\times {\bf L}'\rangle \Big)\bigg]\tau^0\,, \nn  \eea
%
%
%
\bea 2\lambda \langle {\bf L}'\rangle \nabla h \lambda^{-1} &=& \frac{1}{4U^2}\Big(3{\bf L}'\cdot \delta {\bf L}+{\bf L}'\cdot {\bf \Sigma}' \Big)\nn
\\
&& + \tau^0\bigg\langle \frac{1}{2U^2} ( {\bf X}'\cdot \delta {\bf X}+ 2 {\bf X}\cdot \delta {\bf X}'){\bf L}' + \frac{3}{4U^2} \delta {\bf L}  \times{\bf L}' -\frac{1}{4U^2} {\bf L}'\times {\bf \Sigma}'  \bigg\rangle \tau^0\,, \nn
\eea
%
%
%
%
%
\bea
 \lambda (\lambda^{-1}\delta \lambda)''\lambda^{-1} &=& \left(\frac{\delta U}{U}\right)'' + \tau^0\bigg[ \frac{1}{U^2}\langle \delta {\bf L}' \rangle -\frac{2}{U^2}\langle {\bf X}'\times \delta {\bf X}' \rangle  + \frac{4}{U^4}(2U'\delta U-{\bf X}\cdot \delta {\bf X}')\langle {\bf L} \rangle \nn
 \\
 && ~~~~~  -\frac{4U'}{U^3}\langle \delta {\bf X}\times {\bf X}' \rangle  -2\frac{\delta U}{U^3}\langle  {\bf L}'\rangle   + \frac{2}{U^4}({\bf X}\cdot {\bf X}''-UU'' -U'^2)\langle {\bf \Sigma}\rangle \bigg]\tau^0\,,  \nn
 \\
 &=&  \left(\frac{\delta U}{U}\right)'' + \tau^0\bigg[ \frac{1}{U^2}\langle \delta {\bf L}' \rangle +\frac{2}{U^4} \langle {\bf L}\times \delta {\bf L} \rangle  + \frac{1}{U^2}\Big(6\frac{U'\delta U}{U^2}-2\frac{\delta U'}{U}\Big)\langle {\bf L} \rangle   \nn
 \\
 &&~~~~~~~~~~~~~~~~ -2\frac{U'}{U^3}\langle \delta {\bf L} \rangle  -2\frac{\delta U}{U^3}\langle {\bf L}' \rangle\bigg]\tau^0\,, \nn
\eea
\bea
\lambda(\lambda^{-1}\lambda'\lambda^{-1}\delta \lambda)'\lambda^{-1} &=& \left(\frac{U'\delta U}{U^2}\right)'+ \frac{1}{U^4}{\bf L}'\cdot {\bf \Sigma} -2\frac{\delta U}{U^5}{\bf L}^2  -2\frac{U'}{U^5} {\bf L}\cdot {\bf \Sigma} + \frac{1}{U^4}{\bf L}\cdot \delta {\bf L}  \nn
\\ && + \tau^0\bigg[\frac{1}{U^2}\Big(\frac{\delta U'}{U}- 5\frac{U'\delta U}{U^2}\Big) \langle {\bf L} \rangle + \frac{U'}{U^3} \langle \delta {\bf L} \rangle + \frac{\delta U}{U^3} \langle {\bf L}' \rangle +\frac{1}{U^2}\left(\frac{U'}{U}\right)'\langle {\bf \Sigma} \rangle  \nn
\\
&&~~~~~  +2\frac{U'}{U^5}\langle {\bf L} \times {\bf \Sigma} \rangle  - \frac{1}{U^4} \langle {\bf L}' \times {\bf \Sigma} \rangle - \frac{1}{U^4} \langle {\bf L} \times \delta {\bf L} \rangle \bigg]\tau^0\,. \nn
\eea
\bea
\lambda  \left(\frac{\delta U}{U}\lambda^{-1}\lambda'\right)'\lambda^{-1} &=&   \left(\frac{U'\delta U}{U^2}\right)'+  \tau^0\bigg[ \Big(\frac{\delta U'}{U^3}-3\frac{U'\delta U}{U^4}\Big)\langle {\bf L} \rangle  + \frac{\delta U}{U^3}\langle {\bf L}' \rangle \bigg]\tau^0\,.    \nn \eea
%
Note that
\bea    \lambda_{ac}  (\nabla_{[\rho} h^{c}_{d]})' \lambda^{-1\, db} & =&  \half \left(\frac{\delta U}{U}\right)'' + \frac{3}{4}\left(\frac{U'\delta U}{U^2}\right)'  + \frac{1}{4U^4}{\bf L}'\cdot {\bf \Sigma} -\frac{\delta U}{2U^5}{\bf L}^2  -\frac{U'}{2U^5} {\bf L}\cdot {\bf \Sigma} + \frac{1}{4U^4}{\bf L}\cdot \delta {\bf L}  \nn
\\ && +\tau^0\bigg[ \frac{1}{2U^2}\langle   \delta {\bf L}' \rangle  -\frac{1}{4U^2}\left(\frac{\delta U}{U}\right)'   \langle {\bf L} \rangle -\frac{3}{4}\frac{U'}{U^3} \langle \delta {\bf L} \rangle - \frac{1}{4}\frac{\delta U}{U^3} \langle {\bf L}'\rangle \nn
\\&&~~~ + \frac{1}{4U^2}\left(\frac{U'}{U}\right)' \langle {\bf \Sigma} \rangle + \frac{U'}{2U^5} \langle {\bf L}\times {\bf \Sigma} \rangle -\frac{1}{4U^4} \langle {\bf L}'\times {\bf \Sigma} \rangle + \frac{3}{4U^4} \langle {\bf L}\times \delta {\bf L} \rangle \bigg]\tau^0\,, \nn \eea
%
%
%
For a convenience of our calculation, let us set
\[ \lambda_{ac}  (\nabla_{[\rho} h^{c}_{d]})' \lambda^{-1\, db} = H + \tau^0\langle {\bf H} \rangle \tau^0\,,
\]
which leads to
\bea   \lambda'_{ac}  (\nabla_{[\rho} h^{c}_{d]})' \lambda^{-1\, db} & =&\Big[\tau^0\Big(-\frac{U'}{U} + \frac{1}{U2}\langle {\bf L}\rangle\Big)\tau^0  \lambda\Big]_{ac}  (\nabla_{[\rho} h^{c}_{d]})' \lambda^{-1\, db}\,, \nn
\\
&=& \frac{U'}{U}H + \frac{1}{U^2}{\bf L}\cdot {\bf H} + \tau^0\bigg[\frac{U'}{U}\langle {\bf H}\rangle + \frac{1}{U^2}H\langle {\bf L}\rangle - \frac{1}{U^2}\langle {\bf L}\times {\bf H} \rangle\bigg]\tau^0\,,
\eea
where
\bea && U^4\Big[\frac{U'}{U}  {\bf H} + \frac{H}{U^2}  {\bf L}  - \frac{1}{U^2}  {\bf L}\times {\bf H}\Big]  \nn
\\ &=& \frac{UU'}{2} \delta {\bf L}' -\frac{3}{4}\Big(U'^2 + \frac{{\bf L}^2}{U^2}\Big)\delta {\bf L} -\frac{1}{4}\Big(U'\delta U + \frac{1}{U^2}{\bf L}\cdot {\bf \Sigma}\Big){\bf L}' \nn
\\
&& + \Big( U^2 H - \frac{UU'}{4} \Big(\frac{\delta U}{ U}\Big)'   + \frac{U'}{2U^3}{\bf L}\cdot {\bf \Sigma}+ \frac{3}{4U^2}{\bf L}\cdot \delta {\bf L} \Big){\bf L} + \bigg( \frac{UU'}{4}\left(\frac{U'}{U}\right)' - \frac{U'}{2U^3}{\bf L}^2  + \frac{{\bf L}\cdot {\bf L}'}{4U^2} \bigg){\bf \Sigma} \nn
\\
&& -\frac{1}{2}{\bf L}\times \delta {\bf L}' +\frac{\delta U}{4U} {\bf L}\times {\bf L}' +\frac{3U'}{2U} {\bf L}\times \delta {\bf L} - \frac{U'}{4U}{\bf L}'\times {\bf \Sigma} + \frac{1}{U^2}\Big(U'^2 -\frac{1}{4}(UU')'\Big){\bf L}\times {\bf \Sigma}\,. \nn
\eea
%
%
%
Some other useful formulae to obtain $Q^{\mu\nu}_K$:
\bea    (\lambda'\lambda^{-1}\lambda')^{a}_{c}\nabla_{[\rho} h^{c}_{b]}\lambda^{-1\, bt} &=& \frac{U^2U'^2 + {\bf L}^2}{4U^6}({\bf X}'\cdot \delta {\bf X} + 2{\bf X}\cdot \delta {\bf X}') + \frac{U'}{4U^5}{\bf L}\cdot (3\delta {\bf L} +\Sigma') \nn
\\
&& + \tau^0\bigg\langle \frac{U'}{2U^5}({\bf X}'\cdot \delta {\bf X} + 2{\bf X}\cdot \delta {\bf X}'){\bf L} + \frac{1}{8U^6}(U^2U'^2+{\bf L}^2)(3\delta {\bf L} +{\bf \Sigma}') \nn
\\&&~~~~~ - \frac{U'}{4U^5}{\bf L}\times(3\delta {\bf L} + {\bf \Sigma}')\bigg\rangle \tau^0\,, \nn
\eea
\bea  &&  \lambda'_b\Big[  \frac{U'}{U}\nabla_{[\rho}h^{b}_{a]}+\half(\lambda^{-1}\lambda')^{b}_{c}\nabla_{[\rho} h^{c}_{a]} - \half \nabla_{[\rho}h^{b}_{c]}(\lambda^{-1}\lambda')^{c}_{a}\Big]\lambda^{-1\, at} \nn
\\
&=&  \frac{U'^2}{4U^4}({\bf X}'\cdot \delta {\bf X}+2{\bf X}\cdot \delta {\bf X}') + \frac{U'}{8U^5}{\bf L}\cdot (3\delta {\bf L}+{\bf \Sigma}') \nn
\\
&&  + \tau^0\bigg\langle \frac{1}{8U^4}\Big(U'^2+\frac{{\bf L}^2}{U^2} \Big)  (3\delta {\bf L}  +  {\bf \Sigma}' ) +\Big(\frac{U'}{4U^5}({\bf X}'\cdot \delta {\bf X}+2{\bf X}\cdot \delta {\bf X}')   - \frac{1}{8U^6} {\bf L}\cdot  (3\delta {\bf L}  +  {\bf \Sigma}' ) \Big)   {\bf L}  \nn
\\&&~~~~~~~
-\frac{U'}{4U^5}  {\bf L}\times (3\delta {\bf L}  +  {\bf \Sigma}' ) \bigg\rangle\tau^0 \,. \nn
\eea
\bea
\half \lambda'\lambda^{-1}\lambda'[\lambda^{-1}\lambda', \lambda^{-1}\delta \lambda]\lambda^{-1} = -\frac{1}{U^4}\tau^0\bigg\langle \frac{1}{U^2}\Big(U'^2 + \frac{{\bf L}^2}{U^2}\Big){\bf L}\times {\bf \Sigma}+2\frac{U'}{U}\Big(   \frac{{\bf L}\cdot {\bf \Sigma}}{U^2} {\bf L} -\frac{{\bf L}^2}{U^2}{\bf \Sigma} \Big) \bigg\rangle \tau^0\,. \nn \eea

%

Now, we present various formulae to obtain vector expression for
each term in the ADT potential. Since $h=0$ in this formalism, the
relevant terms   in the calculation of $Q_{K}^{\rho t}$ are the
following eleven ones:
\bea
 \nabla^2Q^{\rho t}_R  
 &=&   \frac{\zeta^4}{2} k^{t}\bigg[-U^2({\bf X}\cdot \delta {\bf X}')'' -UU' ({\bf X}\cdot \delta {\bf X}')'  + \frac{(UU')'}{2}   {\bf X}\cdot \delta {\bf X}'  +  {\bf L}\cdot \delta {\bf L}'   +\half {\bf L}'\cdot \delta {\bf L} \bigg] \nn
 \\
 && +  \frac{\zeta^4}{2}\bigg[k\tau^0\bigg\langle U^2  \delta {\bf L}''   + UU' \delta {\bf L}'  - \frac{(UU')' }{2} \delta {\bf L}  - ({\bf X}\cdot \delta {\bf X}')'\, {\bf L}\nn
 \\&& ~~~~~~~~~~~ - \half  ({\bf X}\cdot \delta {\bf X}')  {\bf L}'    +    {\bf L}\times \delta {\bf L}'  -\half \delta {\bf L}\times {\bf L}'   \bigg\rangle \tau^0\bigg]^{t}\,,
 \\
 -2Q^{\alpha[\rho}R^{t]}_{\alpha}  
 &=&\frac{\zeta^4}{4}k^{t}\bigg[\Big(3(UU')'-{\bf X}'^2\Big){\bf X}\cdot \delta {\bf X}' - {\bf L}'\cdot \delta {\bf L}\bigg] \nn
 \\
 && + \frac{\zeta^4}{4}\bigg[k\tau^0\bigg\langle ({\bf X}\cdot \delta {\bf X}') {\bf L}' -\Big(3(UU')'-{\bf X}'^2\Big) \delta {\bf L} + \delta {\bf L}\times {\bf L}'\bigg\rangle \tau^0\bigg]^{t}\,,
 \\
 \frac{1}{8}Q^{\rho t}_{R^2}  
 &=& \frac{\zeta^4}{16}k^t\bigg[ \Big(4(UU')' - {\bf X}'^2\Big) ({\bf X}\cdot \delta {\bf X}')  + 4U^2\Big(2(U\delta U)''' -({\bf X}'\cdot \delta {\bf X}')'\Big)   \nn
 \\&& ~~~~~~~
 -2UU'\Big( 2(U\delta U)'' - ({\bf X}'\cdot \delta {\bf X}')\Big) + 4 U\delta U\Big(2(UU')'' -{\bf X}'\cdot {\bf X}''\Big) \bigg] \nn
 \\
 && - \frac{\zeta^4}{16}\bigg[k\tau^0 \bigg\langle \Big(4(UU')' -{\bf X}'^2\Big) \delta{\bf L}  +2 \Big(2(U\delta U)''-{\bf X}'\cdot \delta {\bf X}' \Big) {\bf L}\bigg\rangle    \tau^0  \bigg]^{t}   \,,
 \eea
 \bea
  && -2\nabla^{\alpha}\xi^{\beta}\nabla_{\alpha}\nabla^{[\rho}h^{t]}_{\beta} \nn
  \\
  &=& -\zeta^4k^{t}\bigg[\half {\bf L}\cdot \delta {\bf L}' + \frac{UU'}{2}(U\delta U)'' - \frac{1}{8}(12U'^2-{\bf X}'^2)(U\delta U)' + \frac{U'}{4U} \Big(4U'^2 - (UU')' +2 {\bf X}'^2  \Big)U\delta U  \nn
  \\ && ~~~~~~~  + \frac{1}{8}(8U'^2-{\bf X}'^2) {\bf X}\cdot \delta {\bf X}'   + \frac{U'}{2U}{\bf L}\cdot \delta {\bf L} +  \frac{U'}{4U}{\bf L}'\cdot {\bf \Sigma} - \frac{\delta U}{4U}{\bf L}\cdot {\bf L}' + \Big(\frac{(UU')'}{4U^2}-\frac{U'^2}{U^2}\Big){\bf L}\cdot {\bf \Sigma}\bigg] \nn
  \\
  && - \zeta^4\bigg[k\tau^0\bigg\langle  \frac{UU'}{2}\delta {\bf L}' -\half {\bf L}\times \delta {\bf L} + \frac{\delta U}{4U}{\bf L}\times {\bf L}' - \frac{U'}{4U}{\bf L}'\times {\bf \Sigma} + \frac{1}{16} \Big({\bf X}'^2-2(UU')' \Big)(\delta {\bf L}  - {\bf \Sigma}' )  \nn
  \\
  &&~~~~~~~~~     - \frac{1}{4}\Big(U'\delta U + \frac{{\bf L}\cdot {\bf \Sigma}}{U^2}\Big){\bf L}'  + \frac{{\bf L} \cdot {\bf L}'}{4U^2} {\bf \Sigma}   + \Big( \half (U \delta U)''  - \half {\bf X}'\cdot \delta {\bf X}'  + \frac{{\bf L}'\cdot {\bf  \Sigma}}{4U^2}\Big){\bf L}
 \bigg\rangle\tau^0\bigg]^{t}\,.
 \eea
  \bea
  -4\xi^{\alpha}R_{\alpha\beta}\nabla^{[\rho}h^{t]\beta}  
 &=& \zeta^4k^{t}\bigg[ \frac{(UU')'}{2}\Big((U\delta U)' +  {\bf X}\cdot \delta {\bf X}'\Big) +\frac{1}{4} {\bf L}'\cdot (3\delta {\bf L} + {\bf \Sigma}')\bigg]
 \\
 &&+\zeta^4\bigg[k\tau^0\bigg\langle   \frac{(UU')'}{4}(3\delta {\bf L} + {\bf \Sigma}') + \half \Big((U\delta U)' +  {\bf X}\cdot \delta {\bf X}'\Big)  {\bf L}' -\frac{1}{4}{\bf L}'\times (3\delta {\bf L} + {\bf \Sigma}' ) \bigg\rangle \tau^0\bigg]^{t}\,,   \nn
 \eea
 \bea
  2\xi^{\alpha}h^{\beta[\rho}\nabla_{\beta}R^{t]}_{\alpha}  
  &=& \zeta^4k^{t}\bigg[ (UU')''\, U\delta U +\frac{1 }{4}({\bf X}\cdot {\bf X}'')\Big( U'\delta U + \frac{1}{U^2} {\bf L} \cdot  {\bf \Sigma} \Big) \nn
  \\
  &&~~~ -\frac{1}{4}\Big(\frac{U'}{U}{\bf L}'\cdot {\bf \Sigma} + \frac{\delta U}{U}{\bf L}\cdot {\bf L}' -\frac{1}{U^2}{\bf L}'\cdot ({\bf L}\times {\bf \Sigma})\Big)\bigg]
  \\
  && +\zeta^4\bigg[k\tau^0 \bigg\langle U\delta U\, {\bf L}'' + \frac{3}{4}\frac{\delta U}{U}{\bf L}\times {\bf L}' -\frac{1}{4}\Big(U'\delta U + \frac{{\bf L}\cdot {\bf \Sigma}}{U^2}\Big) {\bf L}' +\frac{1}{4}\frac{U'}{U}{\bf L}'\times {\bf
  \Sigma} \nn
  \\&&~~~~~~~ -\frac{{\bf X}\cdot {\bf X}''}{8}(\delta {\bf L} -  {\bf \Sigma}')+\frac{1}{4}\Big(2{\bf X}\cdot {\bf X}'' \frac{\delta U}{U} + \frac{{\bf L}'\cdot {\bf \Sigma}}{U^2}\Big) {\bf L} -\frac{{\bf L}\cdot {\bf L}'}{4U^2}{\bf \Sigma}
  \bigg\rangle \tau^0\bigg]^{t}\,,  \nn
 \eea
 \bea
  -Rh_{\alpha}^{[\rho}\nabla^{t]}\xi^{\alpha}  
  &=&
   \frac{\zeta^2}{8}(4{\bf X}\cdot {\bf X}'' + 3{\bf X}'^2)\bigg[ k\,  ({\bf X}'\cdot \delta {\bf X}) + k\tau^0 \Big\langle \half(\delta {\bf L} -  {\bf \Sigma}') \Big\rangle \tau^0\bigg]^{t}
  \,,
  \\
 2 \xi^{[\rho}R^{t]}_{\alpha}\nabla_{\beta}h^{\alpha\beta}   
 &=& -\frac{\zeta^4}{2}k^t (2{\bf X}\cdot {\bf X}'' +{\bf X}'^2) ({\bf X}'\cdot \delta {\bf X} +2{\bf X}\cdot \delta {\bf X}')\,,
 \\
 2\xi_{\alpha}R^{\alpha[\rho}\nabla_{\beta}h^{t]\beta} 
 &=& -\frac{\zeta^4}{2}\Big[ k({\bf X}\cdot {\bf X}')'  + k\tau^0\langle {\bf L}'  \rangle\tau^0 \Big]^{t} ({\bf X}'\cdot \delta {\bf X} +2{\bf X}\cdot \delta {\bf X}')\,,
  \\
 2h^{\alpha\beta}\xi^{[\rho}\nabla_{\alpha}R^{t]}_{\beta}  
 &=& \frac{\zeta^4}{2}k^t\Big[\Big(5{\bf X}'\cdot {\bf X}''-4({\bf X}\cdot {\bf X}')'' \Big) U \delta U - ({\bf X}\cdot {\bf X}'')\, {\bf X}'\cdot \delta {\bf X} +  (UU')\, {\bf X}''\cdot \delta {\bf X}\Big]\,,\nn \\~~~
 \eea
 \bea
  -(\delta R +2R^{\alpha\beta}h_{\alpha\beta})\nabla^{[\rho}\xi^{t]}    &=&\frac{\zeta^4}{2}\Big( (U\delta U)'' +({\bf X}\cdot \delta {\bf X}')'   -{\bf X}''\cdot \delta {\bf X} \Big) \Big[k\, {\bf X}\cdot {\bf X}' + k\tau^0\langle {\bf L} \rangle \tau^0\Big]^{t}\,.~~~~~~~~
 \eea
In the above, we have presented all the preliminary identities. All
we need to do is just combining all the above results and showing
the correction term in the mass formula.

\section{ADT Charges in Schwarzschild Coordinates}
\renewcommand{\theequation}{D.\arabic{equation}}
  \setcounter{equation}{0}
As was explained in the main text,  the linearized  ADT formalism is
insufficient to obtain the correct angular momentum of $WAdS_3$
black holes and  even mass of static new type black holes.    To
obatin the correct result which is consistent with the first law of
black hole thermodynamics and $AdS/CFT$ correspondence, we adopt the
approach augmented by {\it super angular momentum}.  However,
linearized ADT formalism is enough to obtain mass and angular
momentum of BTZ black holes and mass of $WAdS_3$ black holes. In
this appendix, we obtain these charges in Schwarzschild coordinates,
since these coordinates, not ones in Eq.~(\ref{SLmetric}), are
suitable one for $AdS/CFT$ correspondence.

BTZ black holes metric may be written in the form of
\begin{equation}\label{BTZ}
 ds^2 = L^2\bigg[ -\frac{(r^2-r^2_+)(r^2-r^2_-)}{r^2}dt^2
        +\frac{r^2}{(r^2-r^2_+)(r^2-r^2_-)}dr^2  + r^2\Big(d\phi
        + \frac{r_+r_-}{r^2}dt\Big)^2\bigg]\,,
\end{equation}
where the normalization of the time translational and rotational Killing vectors is such that
$\xi_T = \frac{1}{L}\frac{\p}{\p t}, \xi_R =\frac{\p}{\p\phi}$. It is sufficient to take the
next leading term of the metric as $h$-part  for  the ADT potential
\[ h_{tt} =  L^2(r_+^2+r_-^2)\,, \qquad    h_{rr} =    \frac{L^2}{r^4}(r_+^2 + r_-^2)\,, \qquad h_{t\phi} = L^2r_+r_- \,,  \qquad h_{\phi\phi}=0\,. \]
Then, the ADT potentials  are given by
\bea  && Q^{rt}_{R}\Big|_{\xi_T}  = \frac{1}{2L^3r} (r_+^2 + r_-^2)\,,   \qquad Q^{rt}_{K}\Big|_{\xi_T}  =   \frac{1}{4L^5r}(r_+^2 + r_-^2)\,,  \nn \\ &&\nn  \\
&& Q^{rt}_{R}\Big|_{\xi_R} =   \frac{r_+r_-}{L^2r} \,,  \qquad ~~~~~  \qquad Q^{rt}_{K}\Big|_{\xi_R}  =   \frac{r_+ r_-}{2L^4r}\,,   \nn
\eea
which gives us the mass and angular momentum of BTZ black holes in the NMG case as
\beq M = \frac{ ( r_+^2 + r_-^2) }{8G}\eta\Big[ \sigma + \frac{1}{2m^2L^2}\Big]\,, \qquad  J = \frac{ Lr_+ r_-}{4G}\eta\Big[ \sigma + \frac{1}{2m^2L^2}\Big]\,. \eeq
Note that this expression is consistent with the one obtained in various other ways~\cite{Nam:2010dd}.

%
%
%
%

The metric of $WAdS_3$ black holes  in the Schwarzschild coordinates is given by~\cite{Anninos:2008fx}
\begin{equation}
 ds^2 =  L^2 \bigg[-N(r)^2dt^2 +R(r)^2(d\theta +N^{\theta}(r)dt)^2
      + \frac{ dr^2}{4R(r)^2N(r)^2} \bigg]\,,
\end{equation}
where
\begin{eqnarray}
 N(r)^2 &=& \frac{(\nu^2+3)(r-r_+)(r-r_-)}{4R(r)^2}\,,\nn \\
N^{\theta}(r) &=& \frac{2\nu r-\sqrt{r_+ r_- (\nu^2+3)}}{2R(r)^2} \,, \nn \\
R(r)^2 &=& \frac{r}{4} \left( 3(\nu^2-1)r + (\nu^2+3)(r_+ + r_-) -
4\nu\sqrt{r_+ r_-(\nu^2+3)} \right)  \nn
\end{eqnarray}
and $\nu^2 = 3/(4\beta^2-1)$.  \\
It is sufficient to take the metric deviation as
\bea
 h_{rr} &=& \frac{1}{\nu^2+3}\frac{1}{r^3}(r_+ + r_-)\,,  \nn \\
   h_{\theta\theta} &=& \frac{r}{4}\Big[(\nu^2+3)(r_+ + r_-) -4\nu \sqrt{(\nu^2+3)r_+r_-}\Big]\,, \nn \\ h_{t\theta} &=&  -\half \sqrt{(\nu^2+3)r_+ r_-} \,. \nn
\eea
We can also obtain the ADT potentials as
\bea
 Q^{rt}_{R}\Big|_{\xi_T}  &=&    \frac{1}{2L^3} (\nu^2+3)(r_+ + r_-) -2\nu\sqrt{(\nu^2+3)r_+r_-}  \nn \\ && \nn \\
Q^{rt}_{K}\Big|_{\xi_T}   &=&  \frac{1}{4L^5}\bigg[(\nu^2+3)(4\nu^2-3)(r_+ + r_-) - 4\nu(16\nu^2-15)\sqrt{(\nu^2+3)r_+r_-}\bigg]\,.   \nn \eea
These give us mass of $WAdS_3$ black holes solutions of NMG in Schwarzschild coordinates   as
\beq  M = \frac{L^3}{8G}\eta\Big[\sigma Q^{rt}_{R} + \frac{1}{m^2}Q^{rt}_{K} \Big]    = \frac{\nu(\nu^2+3)}{G(20\nu^2-3)}\bigg[\nu(r_+ + r_-) -\sqrt{(\nu^2+3)r_+r_-}\bigg]\,,   \eeq
where we used the relation $1/m^2L^2 = 2\sigma/(3-20\nu^2)$ and
$\eta=\sigma=-1$ in the last equality.  This result is identical
with the one obtained by the $AdS/CFT$
correspondence~\cite{Nam:2010dd}.

\newpage

\end{document}